# Implementing GenAI-Supported Learning in Software Engineering and Computer Science Education using Bloom's Taxonomy

Vəhid Gəruslu[1]
Queen's University Belfast, United Kingdom
v.garousi@qub.ac.uk
Azerbaijan Technical University, Azerbaijan
vahid.garousi@aztu.edu.az

Zafar Jafarov, Aytan Mövsümova, Leyla Məmmədova
Azerbaijan Technical University, Azerbaijan
{zafar.cafarov, aytan.movsumova, leyla.mammadova}@aztu.edu.az

Hüseyn Mirzayev
Qarabağ University, Azerbaijan
huseyn.mirza@qu.edu.az

***Context***: The rapid adoption of generative artificial intelligence (GenAI) tools in software engineering (SE) and computer science (CS) education has introduced new opportunities for learning support, alongside significant concerns related to superficial learning, overreliance, and academic integrity. In response, many institutions have adopted restrictive policies or detection-based approaches that primarily focus on compliance and misconduct. While such measures address short-term risks, they provide limited pedagogical guidance on how GenAI can be meaningfully and responsibly integrated to support learning processes and cognitive development.

***Objective***: This study investigates how explicit instructional guidance aligned with Bloom's taxonomy can support responsible, learning-oriented use of GenAI in SE/CS education. In addition, it explores students' and instructors' perceptions of GenAI-supported learning, including perceived benefits, usage practices, and challenges that arise when GenAI is introduced as a learning supplement rather than a shortcut for task completion.

***Method***: We designed and implemented a Bloom-aligned GenAI usage framework that explicitly articulated appropriate roles for GenAI at different cognitive levels. The framework was embedded into course instructions, laboratory activities, and assessment guidance across multiple SE/CS courses. The study was evaluated using a qualitative, practice-oriented research design, drawing on anonymous open-ended questionnaires and selected learning artifacts. Data were collected from a Software Testing course at Queen's University Belfast and from redesigned undergraduate courses at Azerbaijan Technical University. Responses were analyzed using thematic analysis, with Bloom's taxonomy employed as an analytic lens rather than a measurement instrument. The use of generative AI tools during data analysis was conducted transparently under full human oversight.

***Results***: The findings indicate that students perceived GenAI as most valuable for higher-order cognitive activities, particularly analysis and evaluation, where it supported reasoning, comparison, and reflection. In contrast, GenAI was viewed as less suitable—and sometimes counterproductive—for foundational learning activities such as initial concept acquisition. Explicit Bloom-level guidance influenced students' GenAI usage practices by encouraging reflective interaction, delayed use, and intentional non-use in situations where independent thinking was prioritized. Both students and instructors reported clear pedagogical benefits from the framework, alongside challenges related to increased cognitive effort, the learning curve associated with effective GenAI use, and additional instructional design workload.

---

[1] The other name spelling: Vahid Garousi

***Conclusion***: The study suggests that the educational value of GenAI lies not in unrestricted use or strict prohibition, but in intentional alignment between cognitive learning goals, instructional guidance, and learner self-regulation. Bloom's taxonomy provides a familiar, scalable, and pedagogy-driven framework for operationalizing responsible GenAI use in SE/CS education. The proposed approach offers a practical alternative to enforcement-focused responses and establishes a foundation for future research on cognitively intentional GenAI-supported learning.

**Additional Keywords and Phrases:** Generative AI in education; Software engineering education; Computer science education; Bloom's taxonomy; Responsible AI use; AI-supported learning; Higher-order thinking skills; Instructional design; Academic integrity; Qualitative study

## 1 INTRODUCTION

Generative AI (GenAI) tools are rapidly becoming embedded in software engineering (SE) and computer science (CS) education. Students increasingly rely on these tools for explanations, code generation, debugging, and conceptual clarification. While GenAI offers substantial potential to support learning, reflection, and problem solving, its unstructured adoption has raised concerns related to superficial learning, over-reliance, and academic integrity. In response, many educational institutions have resorted to prohibition or detection-based policies, which address symptoms rather than the underlying pedagogical challenge.

The core problem is not whether students use GenAI, but how they use it. In the absence of explicit instructional guidance, GenAI can accelerate task completion while undermining deep learning, leading to what can be described as cognitive debt: reduced conceptual understanding masked by seemingly correct outputs. Existing approaches provide limited guidance on aligning GenAI use with learning objectives, leaving both students and instructors uncertain about appropriate, responsible, and educationally meaningful use.

As GenAI tools continue to evolve and diffuse across educational contexts, enforcement-focused responses are unlikely to scale or remain effective. Without a pedagogy-driven framework, educators risk either suppressing potentially beneficial learning support or unintentionally encouraging shallow engagement. Addressing this problem is essential to ensure that GenAI contributes to higher-order thinking, supports learner autonomy, and aligns with established educational principles rather than undermining them.

To address this challenge, we design and implement a framework that explicitly aligns GenAI use with Bloom's taxonomy and integrates this guidance into course instructions and learning activities. The framework makes appropriate, delayed, or intentional non-use of GenAI explicit at different cognitive levels and encourages reflective interaction rather than output-driven use. The main research question guiding this study is: *How can Bloom-aligned instructional guidance be embedded in software engineering and computer science education to guide GenAI-supported learning?*

The contributions of this paper are three-fold:

- We present a Bloom-aligned framework that operationalizes responsible GenAI use in SE/CS education (Section 3)
- We provide a qualitative analysis of student and instructor perceptions of GenAI-supported learning across two educational contexts (Section 4)
- We derive practical insights and design implications for educators seeking pedagogy-driven alternatives to prohibition or detection-based approaches (Section 5)

Core take-away messages from the study are as follows:

- GenAI can improve student learning when its use is deliberately aligned with Bloom's cognitive levels and made explicit to students; without such alignment, GenAI risks accelerating task completion for students, while leading to reduced learning and increased "cognitive debt" of students.
- GenAI should be introduced as a learning supplement whose role is explicitly framed per Bloom level and openly discussed with students as part of instructional design.

The remainder of this paper is organized as follows. Section 2 presents the background and related work. Section 3 describes the design and implementation of the Bloom-aligned framework for GenAI-supported learning. Section 4 reports the empirical assessment of the proposed approach. Section 5 discusses key findings, implications, and limitations. Finally, Section 6 concludes the paper and outlines directions for future work.

## 2 BACKGROUND AND RELATED WORK

This section reviews relevant literature on generative AI in education, learning theory, and responsible AI use in software engineering and computer science (SE/CS) education. We first outline Bloom's taxonomy and recent GenAI-integrated extensions, followed by prior work on aligning GenAI with learning objectives in SE/CS contexts. We then discuss conceptual roles of GenAI in learning, acceptable-use and AI maturity models, and research on academic integrity and AI-assisted misconduct. The section concludes by positioning the present study in relation to prior work and clarifying its contribution.

### 2.1 Bloom's Taxonomy and its Recent GenAI-integrated Extension

Bloom's taxonomy is a framework for categorizing educational goals, developed by a committee of educators chaired by Benjamin Bloom in 1956 [1]. The taxonomy has been widely used by educators and also in the education research community worldwide since then.

Figure 1 shows the original 1956 version of Bloom's taxonomy [1] with its revised 2001 revision [2]. While both versions organize cognitive processes hierarchically from lower- to higher-order thinking, the revision modernizes the taxonomy by shifting from noun-based categories (e.g., *Knowledge*, *Comprehension*) to action-oriented verbs (e.g., *Remember*, *Understand*), thereby emphasizing learning as an active cognitive process.

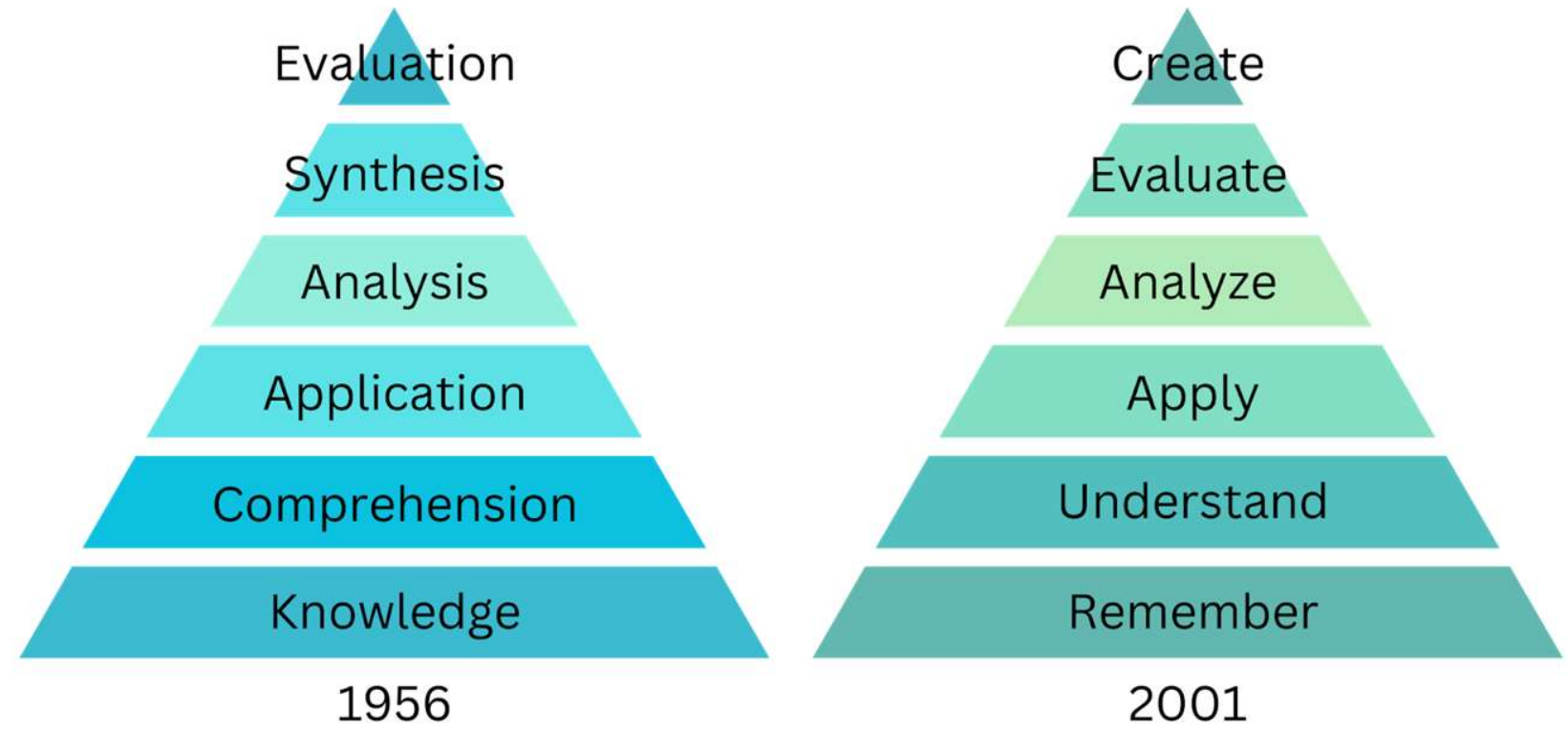

Figure 1-The old version (1956) [1] and latest version (2001) [2] of Bloom's taxonomy

Table 1 presents a GenAI-integrated interpretation of Bloom's taxonomy developed by Oregon State University [3], which explicitly distinguishes between *distinctively human cognitive skills* and the ways in which generative AI can *supplement*—but not replace—learning processes at each cognitive level. The model emphasizes that higher-order activities such as creation, evaluation, and analysis rely fundamentally on human judgment, reflection, ethical reasoning, and lived experience, while GenAI's role is framed as supportive, for example through brainstorming, comparison, critique, and feedback. At lower cognitive levels, GenAI can assist with information retrieval, explanation, and application of known methods, but the framework stresses that core understanding and recall remain human responsibilities.

Table 1-GenAI-integrated Bloom's taxonomy model, designed by the Oregon State University [3]

| Bloom Level | Distinctive Human Skills | How GenAI Can Supplement Learning Processes+ |
|---|---|---|
| 6-Create | Engage in both creative and cognitive processes that leverage human lived experiences, social-emotional interactions, intuition, reflection, and judgment to formulate original solutions. | Support brainstorming processes; suggest a range of alternatives; enumerate potential drawbacks and advantages; describe successful real-world cases; create a tangible deliverable based on human inputs |
| 5-Evaluate | Engage in metacognitive reflection; holistically appraise ethical consequences of alternative courses of action; identify significance or situate within a full historical or disciplinary context | Identify pros and cons of various courses of action; develop and check against evaluation rubrics |
| 4-Analyze | Critically think and reason within the cognitive and affective domains; justify analysis in depth and with clarity | Compare and contrast data, infer trends and themes in a narrowly-defined context; compute; predict; interpret and relate to real-world problems, decisions, and choices |
| 3-Apply | Operate, implement, conduct, execute, experiment, and test in the real world; apply human creativity and imagination to idea and solution development | Make use of a process, model, or method to solve a quantitative or qualitative inquiry; assist students in determining where they went wrong while solving a problem |
| 2-Understand | Contextualize answers within emotional, moral, or ethical considerations; select relevant information; explain significance | Accurately describe a concept in different words; recognize a related example; translate to another language |
| 1-Remember | Recall information in situations where technology is not readily accessible | Retrieve factual information; list possible answers; define a term; construct a basic chronology or timeline |

### 2.2 GenAI and Bloom's Taxonomy in SE/CS Education

Reviewing prior work at the intersection of GenAI and learning objectives in SE and CS education is essential for understanding how GenAI had been previously positioned in relation to student learning. A growing body of research has recently begun to examine the use of GenAI in SE and CS education, with an increasing number of studies exploring its relationship to learning objectives and cognitive skill development. Within this emerging literature, several works have explicitly or implicitly drawn on Bloom's taxonomy to frame GenAI-supported learning activities. In the following, we provide a brief summary of key papers in this area [4-12], focusing on how GenAI has been integrated into SE/CS education, the cognitive levels at which it is typically applied, and the limitations and gaps that remain unaddressed.

An exploratory study by Smith et al. [4] examined early adoption and perceptions of GenAI among CS students prior to the introduction of institutional policies. The findings highlight widespread but unstructured GenAI use and reveal uncertainty about appropriate and educationally meaningful practices, underscoring the need for pedagogical guidance to align GenAI use with learning objectives.

A scoping review [5] synthesized the literature on GenAI–enabled personalization in higher-education computer science and identified conditions under which such personalization supports learning. The review showed that learning benefits are most consistently associated with constrained and scaffolded designs—such as explanation-first guidance, solution withholding, graduated hints, and grounding in student artifacts—rather than unconstrained AI use. While the study does not frame its analysis using Bloom's taxonomy, it highlights the importance of intentional design choices that preserve productive struggle and align AI support with learning processes, reinforcing the need for cognitively guided GenAI integration.

Zastudil et al. [6] reported findings from semi-structured interviews with students and instructors regarding their awareness, experiences, and preferences related to the use of generative AI tools in computing education. The results showed that generative AI was already being used for tasks such as code generation, explanation, and exercise creation, and that both students and instructors expected these tools to play an increasingly significant role in computing classrooms. At the same time, participants expressed concerns about how generative AI should be integrated to effectively support learning goals, and the study highlighted tensions and alignments between student and instructor preferences. The authors discussed implications for curriculum design, assessment practices, and pedagogical approaches, emphasizing the need for intentional integration rather than ad hoc adoption.

Another survey paper [7] analyzed the existing literature on generative AI in CS education using a SWOT framework to examine strengths, weaknesses, opportunities, and threats associated with tools such as ChatGPT. The analysis consolidated early insights into both the capabilities and limitations of generative AI, highlighting concerns related to regulation and plagiarism alongside opportunities for careful and purposeful integration. Based on the synthesis, the authors proposed use cases and guidelines intended to inform future educational strategies and tool development, emphasizing managed adoption rather than outright opposition.

Dickey et al. [8] examined concerns about the impact of generative AI on foundational skill development in computer science students, particularly problem-solving and algorithmic thinking. Based on survey data showing widespread GenAI use, the authors introduced the *AI-Lab* pedagogical framework to guide students' interactions with GenAI in core programming courses. The framework emphasized balanced and reflective use of GenAI through activities such as error identification, explanation, debugging, and prompt formulation, while cautioning against over-dependence that could hinder long-term skill development. The work highlighted the need for instructional guidance that preserves core learning outcomes while leveraging GenAI as a supportive educational tool.

Another survey paper [9] reported a scoping literature review to examine strengths, weaknesses, opportunities, and threats associated with generative AI in computing education following the release of ChatGPT. Analyzing 129 research papers published in 2023–2024 using a SWOT-informed thematic analysis, the author identified challenges such as bias, overreliance, and potential loss of core skills, alongside opportunities related to motivation, pedagogical transformation, and learning support. The study further extended traditional SWOT analysis by proposing a risk management perspective to guide responsible integration of generative AI in computing and higher education contexts.

Sengul et al. [10] reviewed early reactions and emerging research on conversational AI in computing education. The review highlighted polarized responses ranging from defensive, plagiarism-focused approaches to more exploratory perspectives on the role of conversational AI in learning. The findings identified a clear gap between industry practice and higher education in both the pace and scope of adoption, as well as limited empirical research on student experience, teaching practices, and learning tools in software engineering education.

Another paper [11] examined the implications of large language models (LLMs) for SE education and argued that existing research had insufficiently addressed the specific challenges posed by LLMs in this domain. The authors analyzed how LLMs were reshaping SE practice and discussed corresponding implications for educational goals, pedagogy, and ethics. They emphasized the need for a holistic educational approach that combined technical competence, ethical awareness, and adaptive learning strategies to prepare future software engineers for an LLM-rich environment.

Keuning et al. [12] investigated how computing students perceived and used generative AI tools across different courses, programs, and learning contexts over time. Based on three consecutive surveys conducted across a large European research university, the authors analyzed quantitative and qualitative data from students enrolled in a wide range of computing courses with varying programming goals and GenAI usage policies. The findings showed that students' GenAI usage and perceptions varied across course types and evolved over time, highlighting the importance of differentiating and aligning GenAI use with course-specific learning goals rather than treating it uniformly across curricula.

### 2.3 Conceptual Supporting Roles of GenAI in Learning (Tutor, Assistant, Peer)

It is also important to review the prior work on the conceptual roles attributed to generative AI in learning and to understand how GenAI has been framed pedagogically in existing studies. Prior literature variously described GenAI as a tutor, assistant, or peer, shaping expectations about student–AI interaction and levels of learner responsibility [13-17].

A systematic literature review (SLR) [13] synthesized recent empirical studies on human–AI collaboration in STEM education and examined how generative AI was being conceptualized. The findings showed that GenAI was most often used as a tutor or assistant, while emerging approaches increasingly framed it as a collaborative partner supporting higher-order thinking, alongside persistent challenges related to accuracy and pedagogical integration.

Stienstra et al. [14] investigated the use of generative AI as a virtual tutor for teaching a new programming concept to first-year computer science students. Using a ChatGPT-based conversational tutor in a time-constrained learning session, the authors evaluated learning outcomes and student perceptions through post-session quizzes and surveys. The findings suggested that GenAI tutors effectively supported learning and were often preferred for accessibility and convenience, while also revealing risks related to prompt variability and potential overreliance that warranted further pedagogical refinement.

A cross-sectional study [15] examined how preservice teachers in the Global South engaged with generative AI across academic and instructional tasks, based on survey data from teacher education institutions in Ghana. The findings showed that participants used GenAI as a learning companion for accessing explanations and examples, and as a teaching assistant for lesson preparation and assessment-related tasks, with usage increasing in later years of study. While attitudes toward GenAI were generally positive and emphasized support for autonomous learning, challenges such as infrastructural constraints, output inaccuracies, and concerns about academic integrity highlighted the need for explicit GenAI literacy and responsible-use guidance in teacher education programs.

Leon et al. [16] explored the use of generative AI as an on-demand tutoring system to address limitations of traditional education in providing personalized and immediate student support. The authors discussed how GenAI could enable scalable, real-time tutoring through natural language processing, adaptive learning mechanisms, and cloud-based infrastructures, with the aim of reducing learning gaps and inequality. The study outlined future research directions and argued that GenAI-based tutoring had the potential to enhance educational outcomes by offering accessible and personalized learning support at scale.

A mixed-method case study [17] examined the effectiveness of a GenAI-powered Intelligent Tutoring System (named GPTutor) in promoting student engagement among undergraduate students. Survey results showed that interaction with the chatbot component was positively associated with behavioral and emotional engagement, but not with cognitive engagement, while the exercise generator showed no significant associations. Focus-group findings indicated that students used the system selectively based on perceived usefulness, course difficulty, and available support, with usage increasing closer to examinations, and highlighted design improvements—such as enhanced chatbot capabilities and multimodal support—to better align GenAI-based tutoring with learning needs.

### 2.4 Acceptable-Use Models and AI Maturity Frameworks in Education

A number of studies have also been done on acceptable-use models and AI maturity frameworks in education. These approaches help educators and institutions move beyond blanket bans by clarifying what forms of AI use are allowed and how such use should be documented. However, as we review next, these works generally emphasize governance and compliance rather than alignment with learning objectives.

Table 2 illustrates an example of this approach through a five-level acceptable-use model for student AI use in coursework [18], ranging from complete prohibition to full AI use with human oversight. Each level specified the role of AI and corresponding evidence requirements, such as reflections and disclosure of AI interactions. While such models provide practical guidance for managing GenAI use, they do not explicitly connect AI use to cognitive learning goals, leaving open how GenAI could be intentionally aligned with learning processes—an issue addressed by the Bloom-aligned framework proposed in this paper.

Table 2- A 5-level model (scale) for student's acceptable use of AI in coursework [18]

| Level | Description | Evidence / Documentation |
|---|---|---|
| Level 1 – No AI use is allowed | The assessment should be completed with no AI at any time, showing core skills and knowledge. | Invigilated exam or secure in-class work. |
| Level 2 – AI may be used for Planning | The assessment may use AI only for brainstorming, outlining, and early research. Final work must show student refinement. | Reflection, versions of coursework before/after AI use, and chat link/screenshots shall be included in submission. |
| Level 3 – AI may be used for Collaboration | The assessment may use AI for drafting, feedback, and refinement, with critical student editing. | Reflection, versions of coursework before/after AI use, and chat link/screenshots shall be included in submission. |
| Level 4 – AI for Specified Task Completion | AI can be used to complete certain elements of the task, as specified by the teacher. This level requires critical engagement with AI-generated content and evaluating its output. You are responsible for providing human oversight and evaluation of all AI-generated content. | All AI-created content must be cited. Link(s) to AI chat(s) must be submitted with final submission. |
| Level 5 – Full AI Use with Human Oversight | Students may use AI throughout their assessment to support their work in any way they deem necessary. AI should be a "co-pilot" to enhance human creativity. Students are responsible for providing human oversight and evaluation of all AI-generated content. | All AI-created content must be cited. Link(s) to AI chat(s) must be submitted with final submission. |

In addition to acceptable-use scales at the assessment level, prior work also proposed AI maturity models to characterize how educational institutions progressed in their adoption of AI. A UK-based not-for-profit organization, named Joint Information Systems Committee (JISC), presented an AI Maturity Model for Education [19], as shown in Table 3, which presents five stages of organizational adoption of AI, ranging from initial awareness and experimentation to fully embedded and optimized use of AI across institutional processes. The model emphasizes institutional readiness, data maturity, staff skills, governance mechanisms, and the strategic integration of AI into teaching and operational activities.

Table 3- AI Maturity Model for Education (Source: [19])

| **Level 1: Approaching and Understanding** | **Level 2: Experimenting and Exploring** | **Level 3: Operational** | **Level 4: Embedded** | **Level 5: Optimized / Transformed** |
|---|---|---|---|---|
| Interested in AI Understanding how it has impacted or is transforming other sectors | Initial AI guidance produced Experimentation and pilots with existing processes and with existing AI-enabled tools Data culture to support AI emerging Responsible AI processes established | Institutional AI principles established A systematic approach to staff AI skills and literacy Use of everyday AI institution-wide Task-specific AI used for one or more processes across an organization (e.g., chatbots for a specific purpose or adaptive learning systems) | AI embedded in strategy Data maturity allows AI to be considered for all new systems and processes Mature processes to manage the lifecycle of all AI products, including procurement and continuous monitoring | AI is supporting the delivery of learning that optimizes opportunities and outcomes for all learners The right tasks are automated, freeing staff time for creativity and human interaction |

### 2.5 Academic Integrity, AI-assisted Misconduct and "AI-giarism"

The term *"AI-giarism"* has recently begun to appear in academic discourse, e.g., [20, 21], to describe a novel form of plagiarism that specifically involves the use of artificial intelligence tools to generate or reproduce content without appropriate attribution. Scholars have adopted this neologism to highlight how traditional definitions of plagiarism must adapt in response to generative AI technologies: *AI-giarism* encompasses instances in which students present AI-generated material as their own work, challenging established norms of academic integrity and detection practices.

Recent work on academic integrity in the age of generative AI has begun to examine the psychological and institutional factors underlying AI-facilitated plagiarism. For example, Waqas et al. [20] investigated how students' attitudes toward AI and moral disengagement shape engagement in AI-assisted academic misconduct. Based on a large-scale survey of undergraduate students, the study shows that positive attitudes toward AI can mediate the relationship between institutional academic support and AI-giarism, while moral disengagement further moderates this relationship. The authors argue that deterrence-based integrity policies alone are insufficient and propose a mediated model in which institutional context, individual attitudes, and moral reasoning jointly influence student behavior. This work highlights the importance of educational approaches that address not only access to AI tools, but also students' ethical reasoning and understanding of appropriate AI use.

Chan [21] proposed a model that conceptualizes academic misconduct involving generative AI through two parallel spectrums, as shown in Figure 2, contrasting traditional plagiarism with the emerging notion of AI-giarism in the context of students' academic writing.

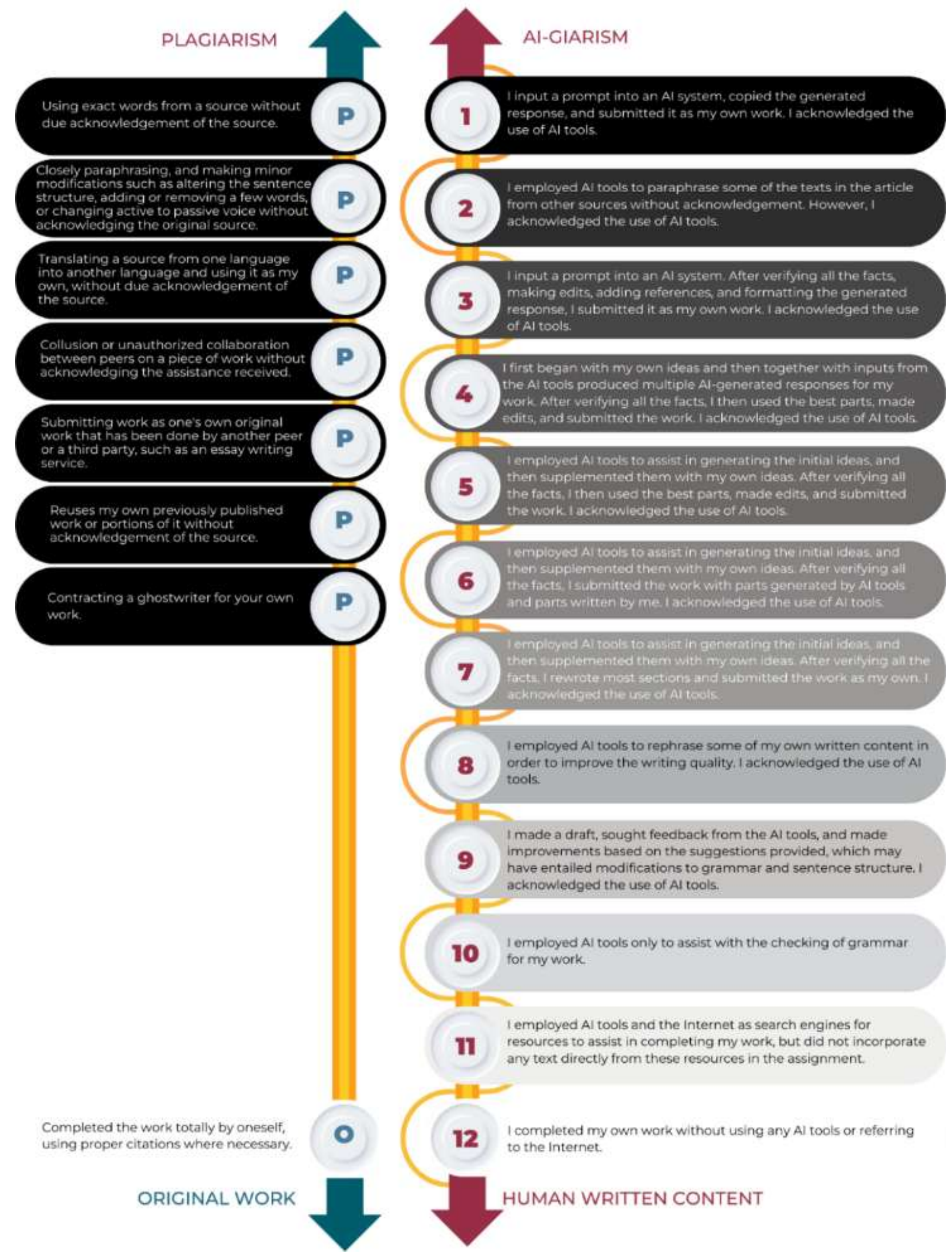

Figure 2- Two spectrums, showing the traditional concept of plagiarism and the new "AI-giarism" concepts in context of students' academic writing, and showing what constitutes ethical or unethical behavior. Source: [21]

On the traditional plagiarism spectrum of Chan [21]'s model (Figure 2), long-established unethical practices—such as verbatim copying, close paraphrasing without attribution, and ghostwriting—are presented as clearly unacceptable. In contrast, the "AI-giarism" spectrum depicts a graded continuum of AI-assisted practices, ranging from overtly unethical behaviors, such as submitting AI-generated content as one's own work, to more acceptable uses, including AI-supported editing, feedback, and grammar checking. By framing AI-assisted writing practices along a continuum rather than as a binary distinction, the model highlights the nuanced roles of student agency, transparency, and intellectual contribution.

Our work builds on this perspective by shifting the focus from identifying misconduct to proactively designing learning activities—grounded in Bloom's taxonomy—that guide students toward ethical, reflective, and pedagogically meaningful uses of GenAI in software engineering and computer science education.

### 2.6 Theoretical Foundations for GenAI-supported Learning

Reviewing theoretical foundations for GenAI-supported learning is relevant for this paper because it helps situate the proposed Bloom-aligned approach within established learning theories and clarify how GenAI-mediated interactions could support—or undermine—cognitive processes central to effective learning.

Social Learning Theory (SLT):

Bandura's Social Learning Theory (SLT) [22] outlines four key elements of how people learn by observing others—especially relevant in today's AI-enhanced learning environments:

- Attention – Learners must notice the modeled behavior, whether from a human instructor or an AI system.
- Retention – They must mentally retain what they observed, often supported by digital content or interactive tools.
- Reproduction – Learners need to apply or simulate the behavior themselves, such as coding or problem-solving.
- Motivation – They must be driven to act, which can come from feedback, gamification, or personal goals.

Paradigms (modes) of AI in Education

A number of studies have characterized different paradigms of AI in education. For example, a paper [23] presented the model shown in Table 4. This model divides AI in education into three paradigms, which conceptualize the evolving roles of AI in learning and instruction. The paradigms range from AI-directed, learner-as-recipient, grounded in behaviorism and exemplified by early intelligent tutoring systems, to AI-supported, learner-as-collaborator, informed by cognitive and social constructivist theories and characterized by interactive and dialog-based systems. The third paradigm, AI-empowered, learner-as-leader, reflects a shift toward learner agency and personalization, drawing on connectivism and complex adaptive systems and emphasizing human–AI cooperation. Together, these paradigms illustrate a progression from AI-driven instruction toward learner-centered, empowerment-oriented educational models, providing a useful theoretical lens for situating contemporary GenAI-supported learning approaches.

Table 4-Different Paradigms (Modes) of AI in Education (Source: [23])

| Paradigm | Theoretical underpinning | Implementations | AI techniques |
|---|---|---|---|
| Paradigm One: AI-directed, learner-as-recipient | Behaviorism | Earlier work of Intelligent Tutoring Systems (ITSs) | AI based on statistical relational techniques |
| Paradigm Two: AI-supported, learner-as-collaborator | Cognitive, social constructivism | Dialogue-based tutoring systems; Exploratory learning environments | Bayesian networks, natural language processing, Markov decision trees |
| Paradigm Three: AI-empowered, learner-as-leader | Connectivism, Complex adaptive system | Human–computer cooperation; Personalized/adaptive learning | Brain–computer interface, machine learning, deep learning |

AI-centric learning models

A number of studies have characterized and presented models for AI-centric learning process. For example, a paper [24] presented a three-level model of learning that explains how AI-supported learning operates in three levels: (1) the individual learner (micro), (2) learning interactions and groups (meso), and (3) the broader educational context (macro).

At the micro level, the model describes learning as a cycle in which students encounter challenges, reflect, adapt their understanding, and gradually build confidence. In this process, AI tools can support learning by providing timely feedback, explanations, and prompts that help learners make sense of new concepts.

At the meso and macro levels, the model shows how AI systems interact with learning artefacts, rules, communities, and institutional practices, shaping how knowledge is shared and supported beyond the individual learner. Overall, the model helps explain how AI can influence learning not only through direct interaction with students, but also by supporting collaboration, instructional design, and educational structures within which learning takes place.

Habits predicting use of AI by students

A number of studies have presented theories and models for factors and habits predicting use of AI by students. For example, a paper [25] applied the Unified Theory of Acceptance and Use of Technology (UTAUT-2) to AI adoption in education. The model explains how factors such as performance expectancy, effort expectancy, social influence, facilitating conditions, hedonic motivation, price value, and habit shape learners' behavioral intention to use AI, which in turn influences their actual use behavior. Importantly, the model highlights that habitual use and perceived usefulness play a central role in sustained AI adoption, sometimes exerting a stronger influence than deliberate intention alone. This perspective is relevant for GenAI-supported learning, as it helps explain why students may adopt AI tools frequently, even in the absence of explicit pedagogical guidance, reinforcing the need for intentional instructional frameworks that shape not just usage, but learning-oriented use.

### *2.7* Importance of Teaching GenAI Literacy as a Core Educational Skill

The rapid diffusion of generative AI across professional and societal domains has made GenAI literacy an essential educational competency rather than an optional technical add-on. As technology leaders such as the CEO of Nvidia Corporation, Jensen Huang, has succinctly stated, "*AI won't replace you. A human using AI will*." This observation captures a central educational challenge: competitive advantage in the age of AI does not stem from mere access to generative technologies, but from the ability to use them effectively, critically, and responsibly. Consequently, higher education institutions cannot assume that students will naturally acquire these skills through ad hoc or unsupervised use of AI tools.

Teaching GenAI literacy involves more than familiarizing students with prompts or interfaces. It requires structured guidance on how to frame problems, evaluate AI-generated outputs, identify limitations and biases, and integrate AI assistance into disciplinary reasoning and decision-making. Without such instruction, students risk either over-reliance on AI systems or superficial use that undermines learning outcomes. Educators therefore play a pivotal role in transforming GenAI from a passive convenience into an active cognitive partner that supports analysis, evaluation, and creative synthesis.

Positioning GenAI literacy as a core educational skill also aligns with broader learning objectives in higher education, including critical thinking, ethical awareness, and lifelong learning. By explicitly teaching students how and when to use generative AI, educators help ensure that AI augments human capability rather than substitutes for it. In this sense, GenAI literacy becomes a foundational competence—analogous to information literacy or programming literacy—necessary for graduates to function effectively in AI-permeated professional environments.

### 2.8 Prior Recent Work Underpinning the Current Study

A recent previous work [26] examined how explicit guidance and transparency could encourage responsible use of generative AI in software engineering and computer science education. That study focused on designing and deploying practical instructional interventions that clarified *when* and *how* students were permitted to use AI tools, with the aim of reducing misuse while maintaining academic integrity. Through empirical investigation, the work highlighted that students were willing to engage responsibly with GenAI when expectations were clearly articulated and when reflection and disclosure were embedded into coursework design.

While the prior study demonstrated the value of explicit guidance for responsible AI use, it primarily addressed GenAI use from a *policy and accountability* perspective rather than from a *cognitively grounded pedagogical* one. In particular, it did not explicitly differentiate GenAI use according to learning objectives or cognitive skill levels, leaving open questions about how GenAI could be intentionally aligned with learning processes rather than merely regulated.

The present study builds directly on that foundation [26] by extending responsible-use guidance into a Bloom-aligned instructional framework. Rather than focusing solely on whether GenAI use is allowed or prohibited, the current work investigates how GenAI can be positioned differently across cognitive levels to support learning-oriented use, reflective engagement, and intentional non-use where appropriate. As such, this study represents a natural progression from governance-focused guidance toward a pedagogy-driven approach for integrating GenAI into SE/CS education.

### 2.9 Positioning and Novelty of the Current Study

Across the reviewed literature, several recurring limitations and gaps emerged that motivated the present study. First, although many studies documented widespread student adoption of GenAI along with perceptions, benefits, and concerns, most focused on describing usage patterns rather than offering actionable pedagogical mechanisms for guiding learning-oriented use. Second, while a number of works proposed guidelines, acceptable-use scales, or high-level frameworks, these approaches were largely policy-driven or task-oriented and did not systematically align GenAI use with cognitive learning goals.

Third, existing studies rarely leveraged established learning frameworks—such as Bloom's taxonomy—to differentiate when and how GenAI should support learning at different cognitive levels. As a result, issues such as cognitive displacement, overreliance, and the erosion of foundational skills were often identified as risks, but seldom addressed through cognitively intentional instructional design. Fourth, much of the available evidence remained cross-sectional, exploratory, or conceptual, with limited empirical examination of how explicit instructional guidance influences students' GenAI usage practices in authentic SE/CS course settings.

Finally, prior work tended to emphasize risks such as plagiarism, overdependence, and skill erosion, while offering comparatively fewer pedagogy-driven alternatives that preserved productive struggle and learner agency while still leveraging GenAI's capabilities. Collectively, these gaps pointed to the need for an empirically grounded, cognitively intentional framework that explicitly aligned GenAI use with learning objectives. The present study addresses this need by proposing and evaluating a Bloom-aligned approach that positions GenAI as a learning supplement whose role varies systematically across cognitive levels.

## 3 THE BLOOM-ALIGNED GENAI-SUPPORTED LEARNING FRAMEWORK AND ITS IMPLEMENTATION IN TWO EDUCATIONAL SETTINGS

This section presents the proposed Bloom-aligned GenAI-supported learning framework together with its concrete implementation in two educational settings. Rather than separating the framework from its practical realization, we intentionally present the conceptual design and implementation side by side to enhance clarity, interpretability, and pedagogical relevance. Given the practice-oriented nature of instructional guidance for GenAI use, the framework is best understood through its instantiation in authentic courses, where abstract principles are translated into explicit teaching materials, learning activities, and assessment guidance. Accordingly, this section first outlines the educational contexts, prior steps, and pedagogical rationale motivating the framework, then introduces the framework itself in an abstract form, and finally demonstrates how it was operationalized in real SE/CS courses. This integrated presentation is intended to make the framework more accessible to educators and to support transfer to similar instructional contexts.

### 3.1 Educational Contexts and Course Settings

This subsection outlines the educational contexts in which the Bloom-aligned GenAI framework was implemented: a Software Testing course at Queen's University Belfast and curriculum updates to Software Engineering and Computer Engineering programs at Azerbaijan Technical University.

#### *3.1.1 Context 1: Software Testing Course in QUB, UK*

The course under study is a final-year undergraduate course titled Software Testing (code name: CSC3056), taught at QUB, and offered once every year. The course has high enrolment numbers, often ranging anywhere between 150-250 students. The course is part of the BSc in Software Engineering and Computer Science program and is designed to equip students with both foundational principles and hands-on experience in software testing. The course syllabus is based on the Certified Tester-Foundation Level (CTFL) certificate of the International Software Testing Qualifications Board (ISTQB); istqb.org. The course has 25 hours of lectures in total, and the list of the lecture topics are shown in Figure 3.

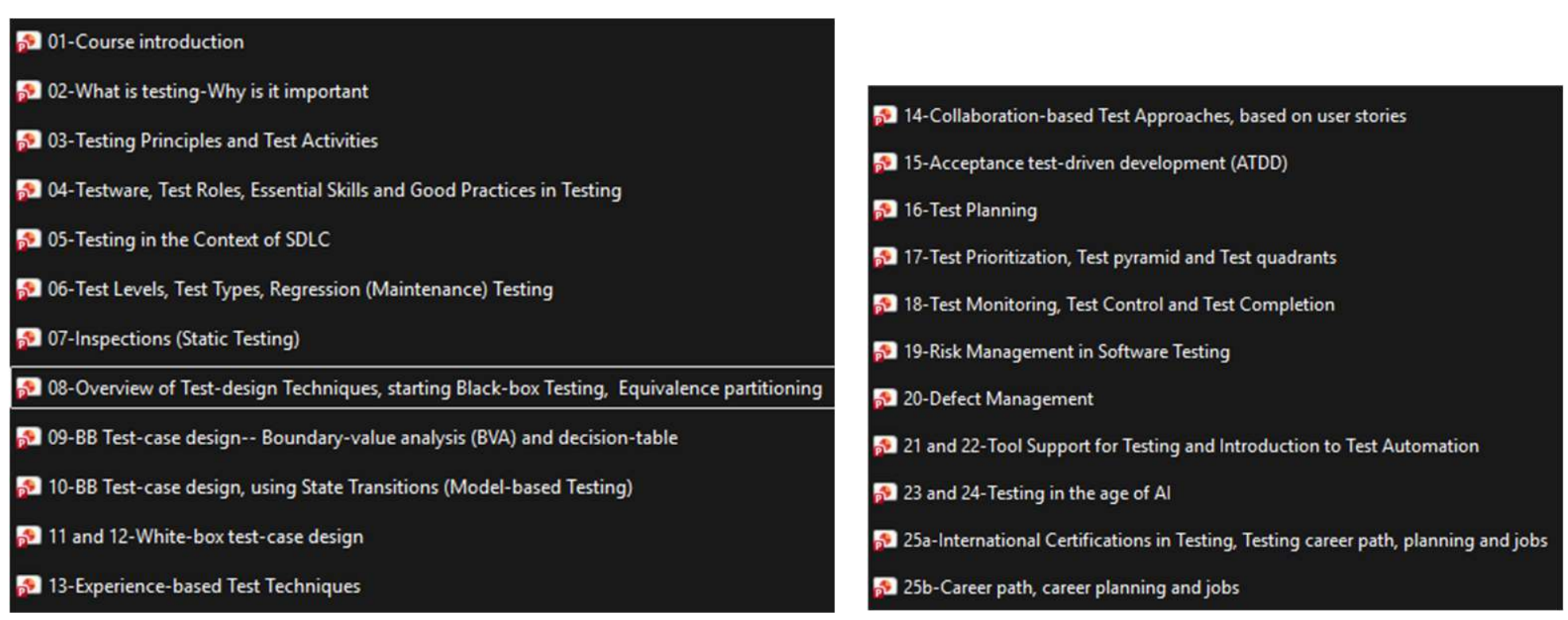

Figure 3-List of topics for each of the 25 lectures of the Software Testing Course

These courses also puts a major emphasis on practical lab exercises, which cover a progression of testing topics, in parallel to the topics taught in the lectures:

- Lab 0: Manual testing and bug tracking
- Lab 1: Java refresher and basic unit testing
- Lab 2: Automated API unit testing using Junit
- Lab 3: White-box testing and code coverage

The labs used in the course are a subset of a widely used open software testing laboratory courseware repository [27, 28]. That repository provides all the lab documents and also the code artifacts needed to conduct the labs. The course emphasizes the development of practical testing skills and critical thinking about software quality. Students work through the full lifecycle of identifying, documenting, and analyzing defects, while learning to design effective test cases and interpret code coverage results. All labs are grounded in the use of industry-relevant test tools and environments to simulate realistic software testing scenarios, e.g., JUnit. This courseware has been used by over 70 instructors in their software-testing courses in more than 15 countries worldwide since 2008 [27].

In recent years, no GenAI-specific instructions were included in the lab instructions. However, like all other educational settings, we have observed that, in recent years, students have started using tools such as ChatGPT to assist with a range of tasks—such as drafting test plans, generating JUnit tests, and paraphrasing bug reports. This spontaneous adoption mirrored global trends in GenAI use across computing education and underscored the need for a reaction and change in the course design, by the course instructor (the first author of the paper), to account for GenAI. We will discuss in Section 3.2 the first step that we took in that direction, by including specific activities in the lab documents for students to actually use GenAI tool, and to encouraging students' responsible use of GenAI.

#### 3.1.2 Context 2: Curriculum update of Software Engineering / Computer Engineering BSc programs in AzTU, Azerbaijan

As the second empirical context, we applied the proposed approach during a phase-by-phase curriculum update of undergraduate courses offered within the Software Engineering, Computer Engineering, Computer Science, and Information Technology BSc programs of the Information Technologies and Telecommunications (ITT) Faculty[1] at Azerbaijan Technical University (AzTU). Unlike context #1, which focused on the redesign of individual courses in isolation, this setting involved a coordinated, faculty-wide course-level intervention across multiple degree programs that share a common pool of core and elective computing courses.

The curriculum update was deliberately conducted course by course, rather than through a disruptive, full-program redesign. This incremental strategy enabled the design team to systematically align existing courses with the pedagogical principles articulated in this paper—particularly the responsible and learning-oriented use of generative AI grounded in Bloom's taxonomy—while preserving curricular stability and accreditation constraints. In each redesigned course, GenAI was explicitly positioned as a supportive learning aid, primarily at higher cognitive levels such as analysis and evaluation, with clear guidance on appropriate use, transparency, and student responsibility.

To date, this approach has been applied to a substantial subset of undergraduate courses that are shared across the ITT Faculty's BSc programs. The courses already redesigned and updated include: Operating Systems, Computer Architecture, Computer Graphics, Computer Networks, Multimedia Technologies, Programming Fundamentals (Python), and Advanced Programming.

---

1 www.aztu.edu.az/sub_site/en/information-and-communication-technologies-13

The remaining courses of the ITT faculty are scheduled for redesign and alignment in the subsequent semester, completing the faculty-level rollout.

### 3.2 Prior Steps Taken Toward Responsible GenAI Use in Coursework

In our prior recent work [26], we took an initial and necessary step toward integrating GenAI in both contexts:. In Context 1 (QUB, UK), we revised the laboratory instructions and coursework documents in an undergraduate software testing course to include concrete, GenAI-related steps. Students were explicitly asked to use GenAI for specific learning tasks and to document, reflect on, and critically review the generated outputs, rather than accepting them uncritically. This made GenAI use explicit, observable, and discussable within routine learning activities..

For Context 2 (AzTU, Azerbaijan), we incorporated considerations for GenAI use during the design of the country's first Software Engineering BSc program. At the program-design stage, stakeholders—including instructors, program coordinators, prospective students, and industry partners—highlighted the importance of establishing clear expectations for responsible GenAI use from the outset. As a result, GenAI was framed as a legitimate learning support tool within the curriculum, accompanied by explicit guidance emphasizing transparency, reflection, and human responsibility in learning and assessment activities.

Overall, our prior recent work [26] primarily addressed *whether* and *how* GenAI can be introduced responsibly into coursework through explicit tasks, transparency, and reflection. That work deliberately focused on mitigating risks associated with unstructured or hidden GenAI use, particularly concerns that GenAI might undermine learning if used uncritically.

At the same time, the earlier study also revealed an important insight: when GenAI use is made explicit and guided, it does not necessarily diminish learning and may, in fact, support it. However, those initial steps did not yet provide a systematic mechanism for guiding *when* and *for what cognitive purposes* GenAI should be used across learning activities. In particular, the prior approach did not distinguish GenAI use based on the cognitive demands of tasks, leaving open the question of how GenAI might move beyond merely being "used responsibly" to actively supplementing student learning.

### 3.3 Need and Rationale for a Bloom-Aligned GenAI Teaching Approach

The limitation of the prior work lies not in the responsible use of GenAI, but in the lack of guidance on how GenAI use should relate to the cognitive demands of learning tasks. While explicit and reflective GenAI use addresses important transparency concerns, it does not ensure that GenAI use aligns with learning goals or supports learning at appropriate levels of cognitive complexity.

Learning activities differ substantially in the type of cognitive engagement they require, ranging from recall and understanding to analysis, evaluation, and creation. Without an explicit mechanism to distinguish these demands, GenAI risks being used uniformly across tasks, regardless of whether such use supports or displaces essential cognitive work.

Bloom's taxonomy was therefore selected as the organizing principle for structuring GenAI use, as it provides a well-established framework for categorizing learning activities by cognitive level. By aligning GenAI use with Bloom's levels, instructors can clarify the intended role of GenAI for different types of tasks and guide students toward level-appropriate use.

In this study, Bloom's taxonomy is adopted as a cognitive structuring mechanism that enables GenAI to supplement learning in a deliberate and task-appropriate manner. The following sections empirically examine how this Bloom-aligned approach is reflected in students' reported learning support and GenAI usage practices.

### 3.4 The Bloom-Aligned GenAI-Supported Learning Framework

This subsection presents the Bloom-aligned GenAI-supported learning framework as a pedagogy-driven instructional design framework for integrating generative AI into SE and CS education. The subsequent subsections then demonstrate how this framework was instantiated in two educational contexts.

The framework builds on prior work that articulates how GenAI can supplement learning processes across Bloom's cognitive levels, such as the extended Bloom model proposed by Oregon State University [3] (as reviewed in Section 2.1), but moves beyond capability-based mappings.

Specifically, while prior models [3] emphasize *what* GenAI can support at each cognitive level, the framework proposed in this study focuses on *how, when, and under what instructional framing* students should interact with GenAI to support learning. The framework therefore emphasizes explicit instructional guidance, learner engagement practices, and pedagogical intent rather than tool capabilities alone.

At the core of the framework are a set of interrelated design elements that together guide learning-oriented and responsible GenAI-use:

- **Cognitive Intentionality:** GenAI use is explicitly aligned with the intended cognitive level of a learning activity, as defined by Bloom's taxonomy. Rather than treating GenAI as a generic productivity aid, instructors clarify the learning intent of each task and articulate how GenAI may or may not contribute to achieving that intent. This makes the cognitive purpose of GenAI use visible to students and helps prevent unreflective reliance that bypasses learning objectives.
- **Temporal Structuring of GenAI Use:** The framework emphasizes when GenAI interaction is pedagogically appropriate within a learning process. Guidance distinguishes between phases such as independent problem exploration, reflective analysis, and post-task evaluation. For example, students may be encouraged to attempt a task independently before engaging GenAI for reflection or comparison, thereby preserving productive struggle while still benefiting from AI-supported feedback.
- **Interaction Modality:** Students are guided on how to interact with GenAI in learning-oriented ways. Instead of requesting direct solutions, they are encouraged to engage GenAI through prompts that support explanation, critique, comparison, or error analysis. This shifts GenAI use from answer generation toward dialogic interaction that supports reasoning and understanding.
- **Prompting (prompt engineering) as a Learning Skill:** Prompt formulation is treated as an explicit component of the learning activity rather than an incidental or hidden practice. Students are guided to construct prompts that reflect learning goals, articulate assumptions, and request specific forms of support. In this way, prompting becomes a transferable learning skill that reinforces metacognitive awareness and intentional engagement with GenAI.
- **Instructor–Student Meta-Dialogue:** Finally, the framework embeds explicit discussion of GenAI use within teaching practice. Instructors and students engage in ongoing dialogue about appropriate use, limitations, and learning implications of GenAI, moving these conversations beyond policy documents into the learning process itself. This shared understanding supports transparency, trust, and learner autonomy.

When combined, the above elements define a pedagogy-driven instructional design framework that operationalizes Bloom-aligned-GenAI-use without reducing learning to tool usage rules. The following sections demonstrate how this framework was instantiated in lectures, laboratory activities, and coursework across two educational settings.

### 3.5 Context #1 Implementation: Integrating Bloom Levels and GenAI into Software Testing Coursework

This subsection describes how Bloom's cognitive levels were explicitly integrated with GenAI guidance in the Software Testing coursework. It explains how the framework was operationalized through learning materials, assessment instructions, and laboratory activities to make appropriate GenAI use visible and intentional at different cognitive levels.

#### *3.5.1 Lecture Materials including Explicit Bloom-Aligned GenAI Guidance and Worked Examples in*

Lecture materials were revised to include worked examples that explicitly demonstrate how GenAI can be used as a learning supplement at specific Bloom cognitive levels. Rather than presenting GenAI as a general-purpose problem solver, the lectures framed GenAI as a tutor or mentor whose role is to support students' reasoning, critique, and reflection, while keeping core problem-solving responsibility with the student.

Figure 4 illustrates a worked lecture example based on the *NextDate* system under test. In this example, students first design equivalence partitions for the day, month, and year inputs manually, without GenAI assistance. Only after completing this task are students instructed to engage GenAI at the Analyze and Evaluate levels of Bloom's taxonomy. GenAI is used to critique the student-designed partitions by highlighting implicit assumptions, potential omissions, and areas requiring closer inspection. Importantly, GenAI is not asked to generate partitions from scratch or to judge correctness in a binary manner. Instead, it acts as a tutor that prompts deeper analysis and supports evaluative reasoning about test adequacy and fault-detection potential. This example makes explicit to students that GenAI's value lies in challenging and refining their thinking rather than replacing it.

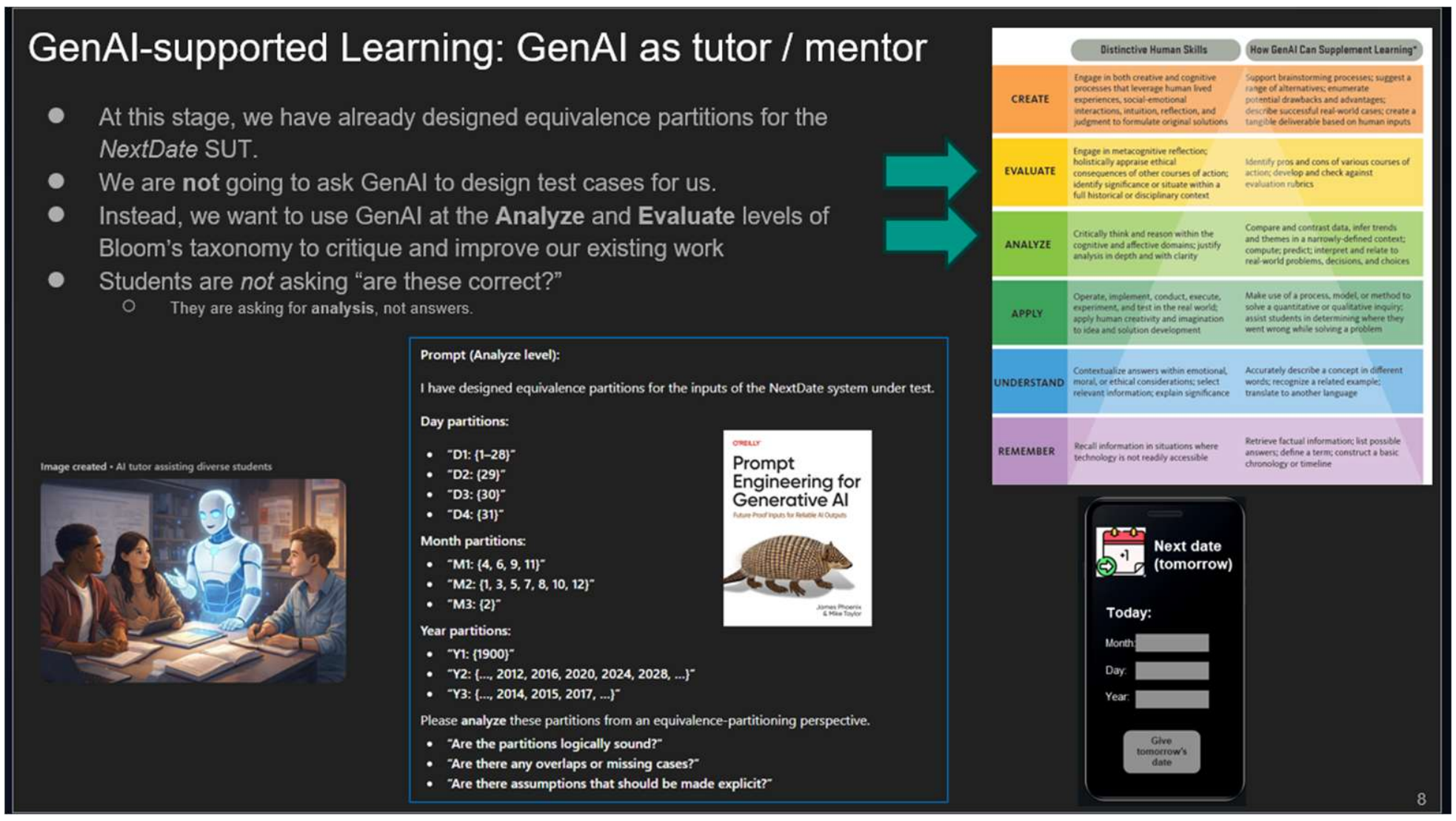

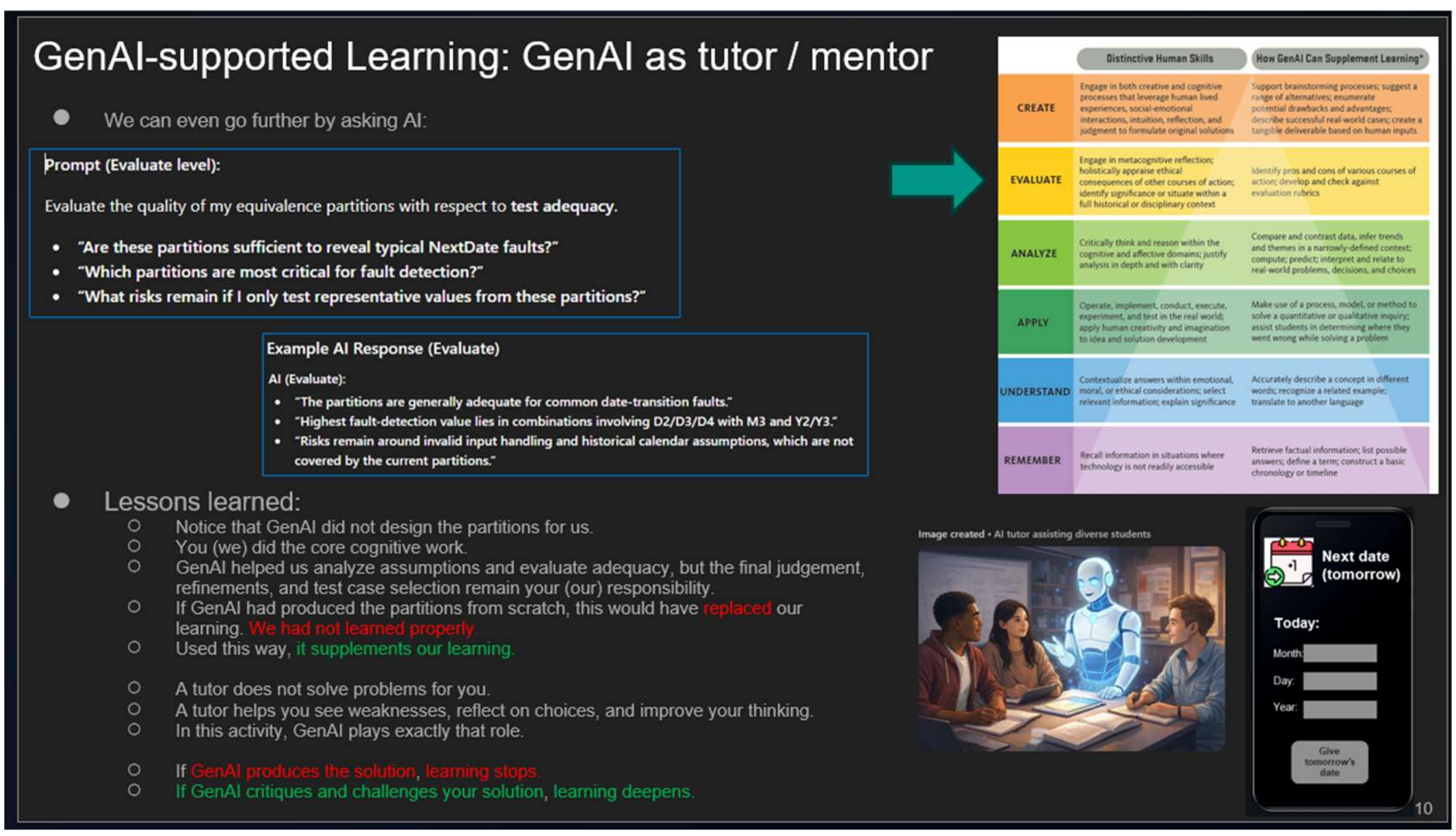

Figure 4-Examples of encouraging GenAI-supported learning in lectures, and using GenAI as a tutor / mentor. Bloom level = Analyze / Evaluate

Figure 5 presents a second lecture example focused on n-way (pairwise and higher-order) testing, a topic already introduced conceptually earlier in the course. Here, GenAI is used primarily at the Apply level of Bloom's taxonomy. Students are shown how GenAI can support the application of a known testing technique to a new and realistic use case—posting a tweet on a social media platform—by helping identify relevant parameters and generate candidate n-way combinations. The lecture then extends the example by encouraging students to move into the Analyze level, where they inspect the generated combinations for realism, missing constraints, and relevance to fault detection. GenAI is again framed as a mentor that explains its reasoning and surfaces issues for discussion, rather than as a tool that produces a final test suite.

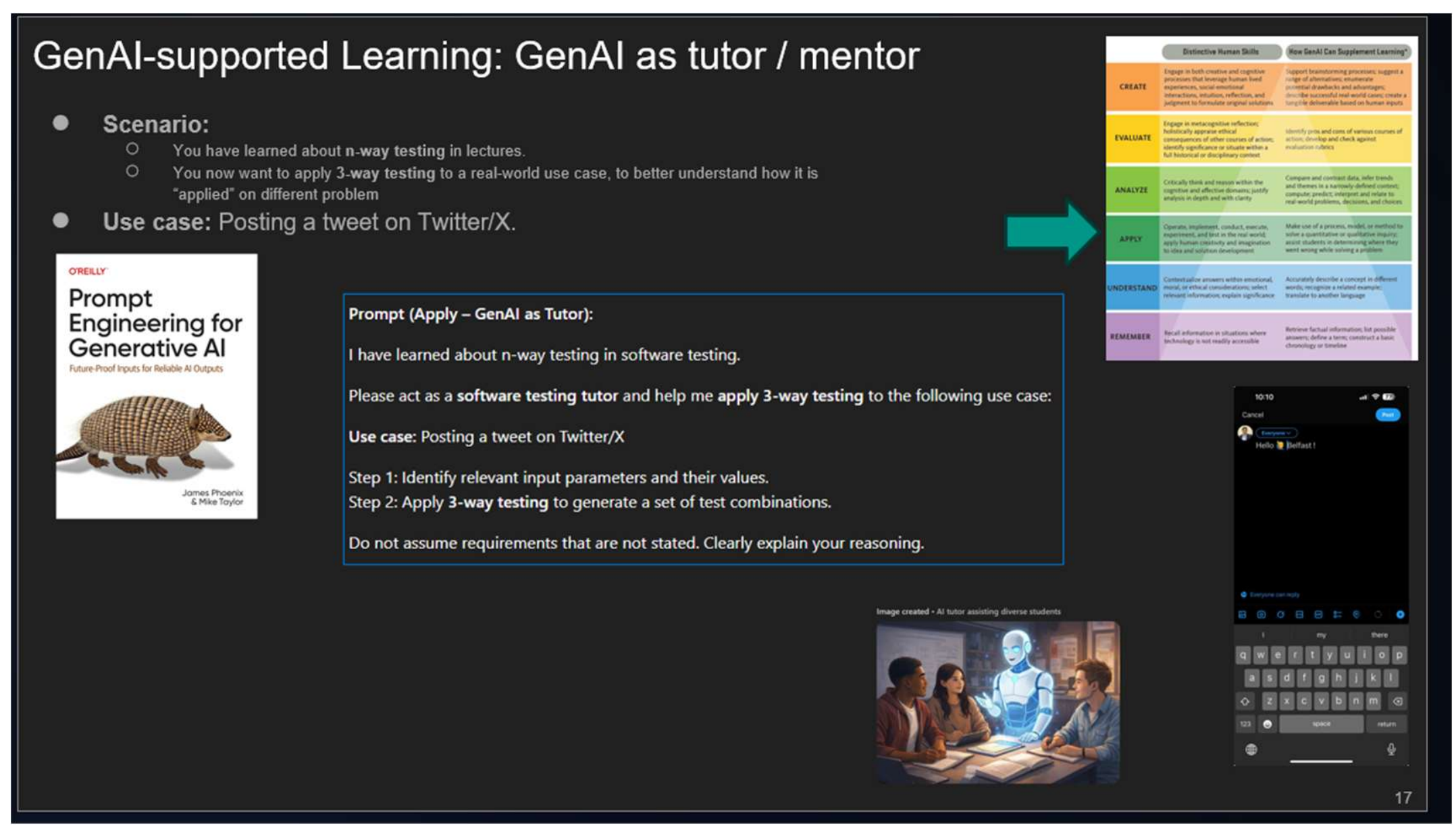

Figure 5-An example of encouraging GenAI-supported learning in lectures, and using GenAI as a tutor / mentor. Bloom level = Apply

In both examples, the lectures explicitly emphasize the intended Bloom level at which GenAI is being used and the corresponding expectations of student engagement. By making this alignment visible in worked examples, students are encouraged to internalize the idea that GenAI's role varies across cognitive levels—from supporting application and exploration to facilitating analysis and evaluation—while core creative and judgement-intensive activities remain student-led.

### *3.5.2 Laboratory Instruction Documents including Explicit Bloom-Aligned GenAI Guidance*

In addition to lecture materials, Bloom-aligned GenAI guidance was operationalized directly within laboratory instruction documents to ensure that expectations regarding GenAI use were explicit, consistent, and visible at the point of task execution. Figure 6 illustrates an example taken from *Lab 0 – Manual Testing and Bug Tracking*, where students are provided with step-by-step guidance on how to use GenAI in a manner aligned with Bloom's cognitive levels.

As shown in the figure, students are first explicitly directed to review a Bloom-aligned model that highlights how GenAI can supplement learning at different cognitive levels. This serves to foreground Bloom's taxonomy as a shared reference framework before any GenAI interaction takes place. Rather than assuming that students will implicitly infer appropriate GenAI usage, the lab document makes the cognitive framing explicit and requires students to consciously situate their GenAI use within specific Bloom levels (e.g., Analyze and Evaluate).

The lab instructions further clarify that, across all GenAI-assisted steps, GenAI is not intended to replace core testing activities such as exploratory test design or judgement-based decision making. Instead, GenAI is positioned as an *additional input* that supports higher-level cognitive activities, including critical inspection of artefacts, evaluation of alternatives, and reflective reasoning. This framing reinforces the role of GenAI as a tutor or mentor that challenges and supports student thinking, rather than as an automated solution generator.

Students are also instructed to make deliberate choices about *how* they interact with GenAI. Specifically, they are guided to avoid prompts that request final or complete answers and to prefer prompts that elicit suggestions, perspectives, or potential issues that can be inspected and assessed. Responsibility for all testing decisions, interpretations, and conclusions is explicitly retained by the student and reinforced as part of the lab reporting requirements.

By embedding this Bloom-aligned GenAI guidance directly into laboratory documents, the approach ensures that GenAI use is not treated as an optional or peripheral activity, but as a structured component of the learning process. The lab instructions thus operationalize the Bloom-aligned framework in a concrete and repeatable manner, making expectations around GenAI use explicit at the point where students engage most actively with testing tasks.

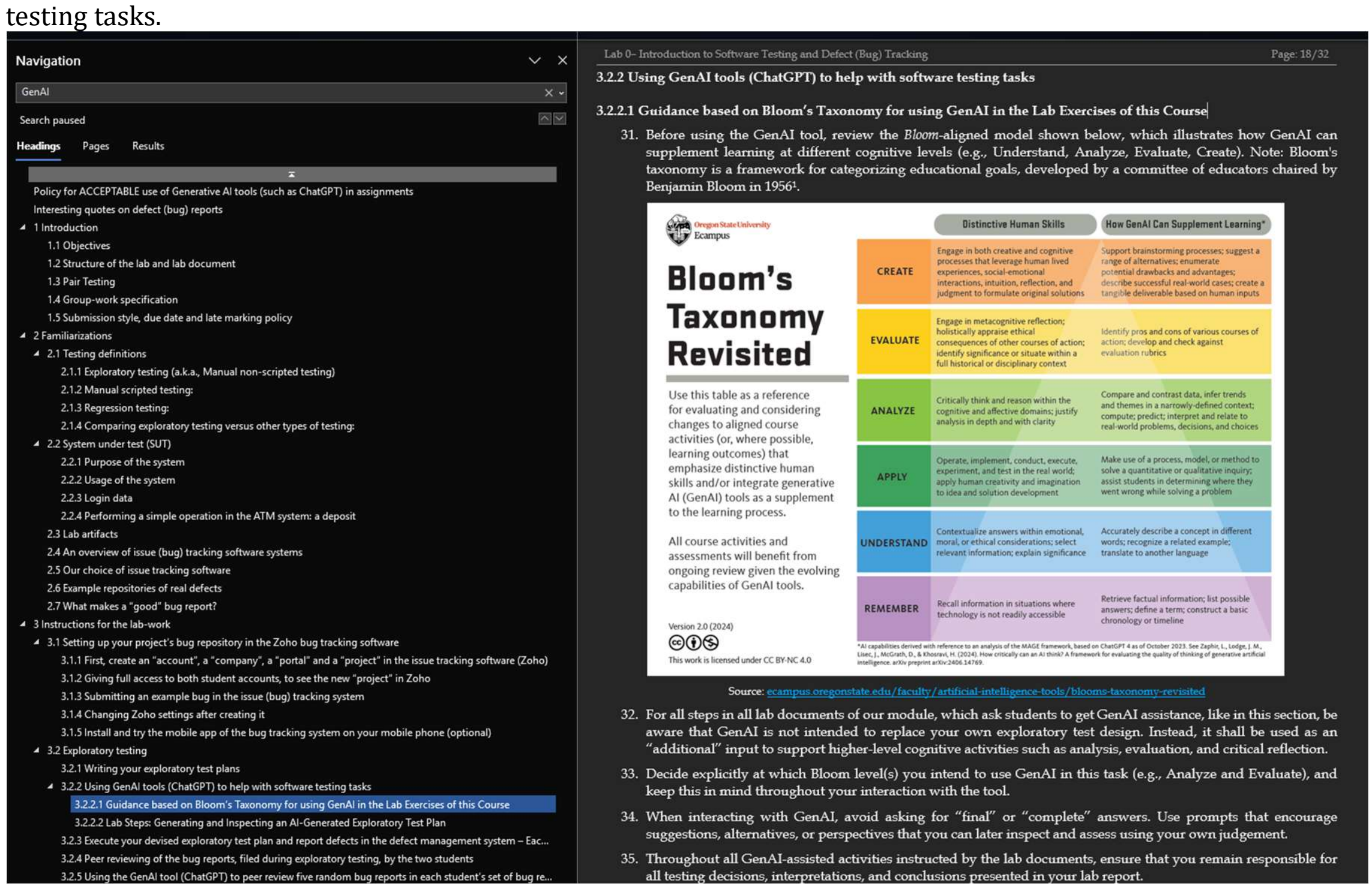

Figure 6-An example of including explicit Bloom-aligned GenAI-usage guidance in laboratory instruction documents. This example is from "*Lab 0-Manual testing and bug tracking*" lab document

In addition to Lab 0, similar Bloom-aligned GenAI guidance was embedded in subsequent laboratory exercises, including *Lab 1 (Java refresher, TDD, and unit testing)*. As shown in the lab1 document screenshot of Figure 7, in lab1, students were explicitly instructed to use GenAI as a tutor or mentor rather than as a code-generation tool. The lab documentation framed GenAI use primarily at the Analyze and Evaluate levels of Bloom's taxonomy, for example when analyzing test logic, evaluating refactoring decisions, or critiquing the adequacy of test cases.

Students were required to decide explicitly which Bloom level they were targeting before interacting with GenAI, to avoid requesting final or complete solutions, and to critically assess the relevance and correctness of any feedback received. Responsibility for all design, coding, and testing decisions remained explicitly with the student, and GenAI interactions had to be documented and reflected upon in the lab report. This ensured consistency with the Bloom-aligned framework while demonstrating how GenAI can supplement learning across multiple labs without displacing core cognitive work.

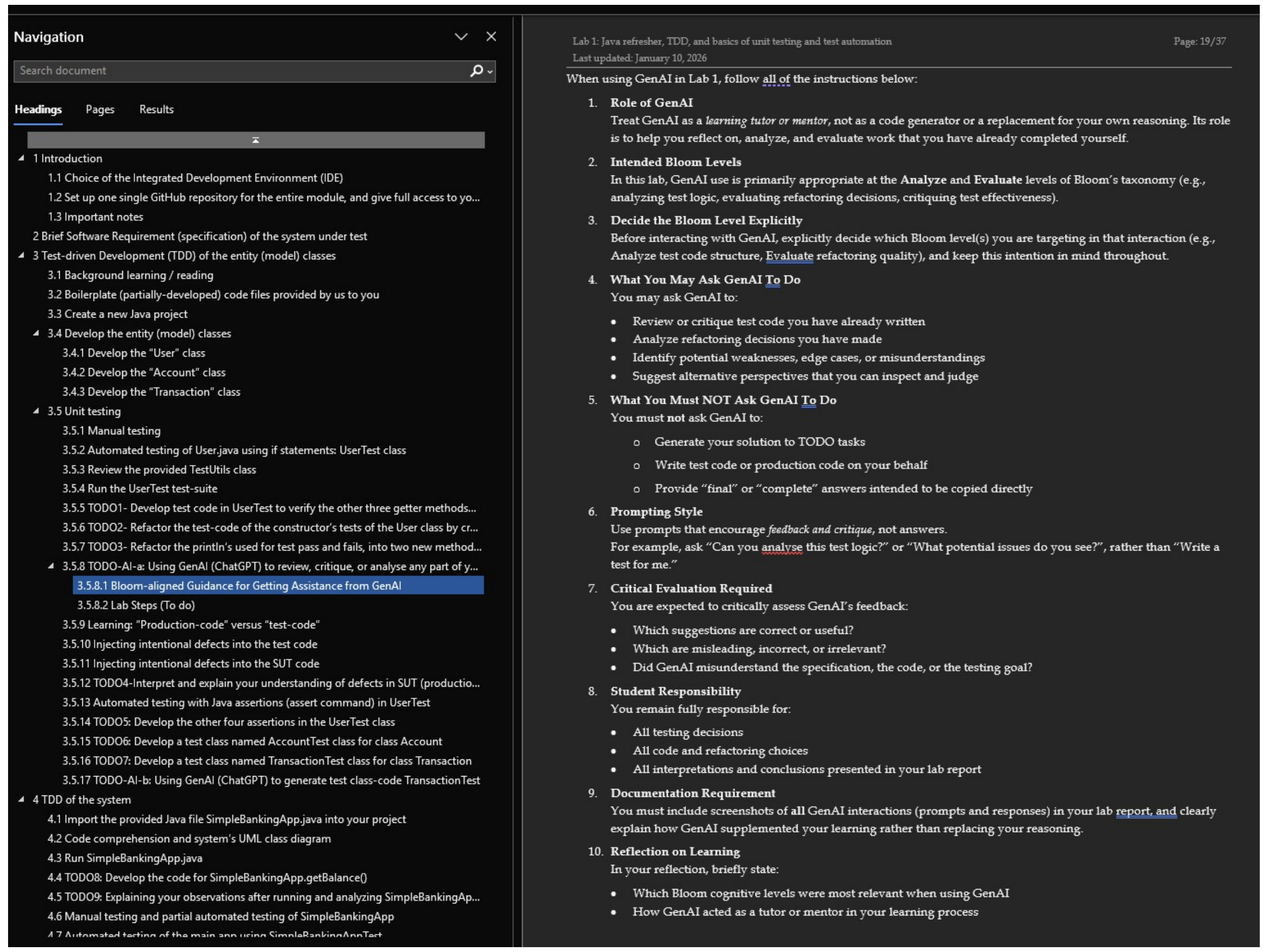
Navigation

Search document

Headings Pages Results

1 Introduction
1.1 Choice of the Integrated Development Environment (IDE)
1.2 Set up one single GitHub repository for the entire module, and give full access to yo...
1.3 Important notes
2 Brief Software Requirement (specification) of the system under test
3 Test-driven Development (TDD) of the entity (model) classes
3.1 Background learning / reading
3.2 Boilerplate (partially-developed) code files provided by us to you
3.3 Create a new Java project
3.4 Develop the entity (model) classes
3.4.1 Develop the "User" class
3.4.2 Develop the "Account" class
3.4.3 Develop the "Transaction" class
3.5 Unit testing
3.5.1 Manual testing
3.5.2 Automated testing of User.java using if statements: UserTest class
3.5.3 Review the provided TestUtils class
3.5.4 Run the UserTest test-suite
3.5.5 TODO1- Develop test code in UserTest to verify the other three getter methods...
3.5.6 TODO2- Refactor the test-code of the constructor's tests of the User class by cr...
3.5.7 TODO3- Refactor the println's used for test pass and fails, into two new method...
3.5.8 TODO-AI-a: Using GenAI (ChatGPT) to review, critique, or analyse any part of y...
3.5.8.1 Bloom-aligned Guidance for Getting Assistance from GenAI
3.5.8.2 Lab Steps (To do)
3.5.9 Learning: "Production-code" versus "test-code"
3.5.10 Injecting intentional defects into the test code
3.5.11 Injecting intentional defects into the SUT code
3.5.12 TODO4-Interpret and explain your understanding of defects in SUT (productio...
3.5.13 Automated testing with Java assertions (assert command) in UserTest
3.5.14 TODO5: Develop the other four assertions in the UserTest class
3.5.15 TODO6: Develop a test class named AccountTest class for class Account
3.5.16 TODO7: Develop a test class named TransactionTest class for class Transaction
3.5.17 TODO-AI-b: Using GenAI (ChatGPT) to generate test class-code TransactionTest
4 TDD of the system
4.1 Import the provided Java file SimpleBankingApp.java into your project
4.2 Code comprehension and system's UML class diagram
4.3 Run SimpleBankingApp.java
4.4 TODO8: Develop the code for SimpleBankingApp.getBalance()
4.5 TODO9: Explaining your observations after running and analyzing SimpleBankingAp...
4.6 Manual testing and partial automated testing of SimpleBankingApp

Lab 1: Java refresher, TDD, and basics of unit testing and test automation — Page: 19/37
Last updated: January 10, 2026

When using GenAI in Lab 1, follow all of the instructions below:

1. **Role of GenAI**
   Treat GenAI as a *learning tutor or mentor*, not as a code generator or a replacement for your own reasoning. Its role is to help you reflect on, analyze, and evaluate work that you have already completed yourself.
2. **Intended Bloom Levels**
   In this lab, GenAI use is primarily appropriate at the **Analyze** and **Evaluate** levels of Bloom's taxonomy (e.g., analyzing test logic, evaluating refactoring decisions, critiquing test effectiveness).
3. **Decide the Bloom Level Explicitly**
   Before interacting with GenAI, explicitly decide which Bloom level(s) you are targeting in that interaction (e.g., Analyze test code structure, Evaluate refactoring quality), and keep this intention in mind throughout.
4. **What You May Ask GenAI To Do**
   You may ask GenAI to:
   - Review or critique test code you have already written
   - Analyze refactoring decisions you have made
   - Identify potential weaknesses, edge cases, or misunderstandings
   - Suggest alternative perspectives that you can inspect and judge
5. **What You Must NOT Ask GenAI To Do**
   You must **not** ask GenAI to:
   - Generate your solution to TODO tasks
   - Write test code or production code on your behalf
   - Provide "final" or "complete" answers intended to be copied directly
6. **Prompting Style**
   Use prompts that encourage *feedback and critique*, not answers.
   For example, ask "Can you analyse this test logic?" or "What potential issues do you see?", rather than "Write a test for me."
7. **Critical Evaluation Required**
   You are expected to critically assess GenAI's feedback:
   - Which suggestions are correct or useful?
   - Which are misleading, incorrect, or irrelevant?
   - Did GenAI misunderstand the specification, the code, or the testing goal?
8. **Student Responsibility**
   You remain fully responsible for:
   - All testing decisions
   - All code and refactoring choices
   - All interpretations and conclusions presented in your lab report
9. **Documentation Requirement**
   You must include screenshots of **all** GenAI interactions (prompts and responses) in your lab report, and clearly explain how GenAI supplemented your learning rather than replacing your reasoning.
10. **Reflection on Learning**
    In your reflection, briefly state:
    - Which Bloom cognitive levels were most relevant when using GenAI
    - How GenAI acted as a tutor or mentor in your learning process

Figure 7-An example of including explicit Bloom-aligned GenAI-usage guidance in laboratory instruction documents. This example is from "*Lab 1- Java refresher, TDD*" lab document

We furthermore provide an example from the "*Lab 2 - Automated API Unit Testing*" lab document, in Figure 8. In Lab 2, which focuses on automated API unit testing, GenAI support was deliberately constrained to activities that align with higher-value cognitive engagement rather than solution generation.

Permitted uses of GenAI for Lab 2 work were limited to tasks such as explaining or paraphrasing Javadoc specifications based strictly on the provided text, analyzing ambiguities or under-specification in requirements, critiquing test-case designs or test code already produced by the student, and identifying potential weaknesses or missing cases. These uses position GenAI as a *tutor or mentor* that supports understanding, analysis, and evaluation, while preserving the student's responsibility for interpreting specifications and designing tests.

Conversely, for Lab 2 work, GenAI was explicitly prohibited from inventing behaviour not stated in the specification or generating complete test artefacts for direct submission. This distinction was intentional: allowing GenAI to assist with interpretation, critique, and reflection can supplement learning by sharpening students' reasoning about requirements and test adequacy, whereas allowing it to generate final solutions would risk displacing essential cognitive work. By clearly separating permitted and prohibited uses in this manner, the lab design operationalizes the Bloom-aligned principle that GenAI should support *how students think*, not replace *what they are expected to do*.

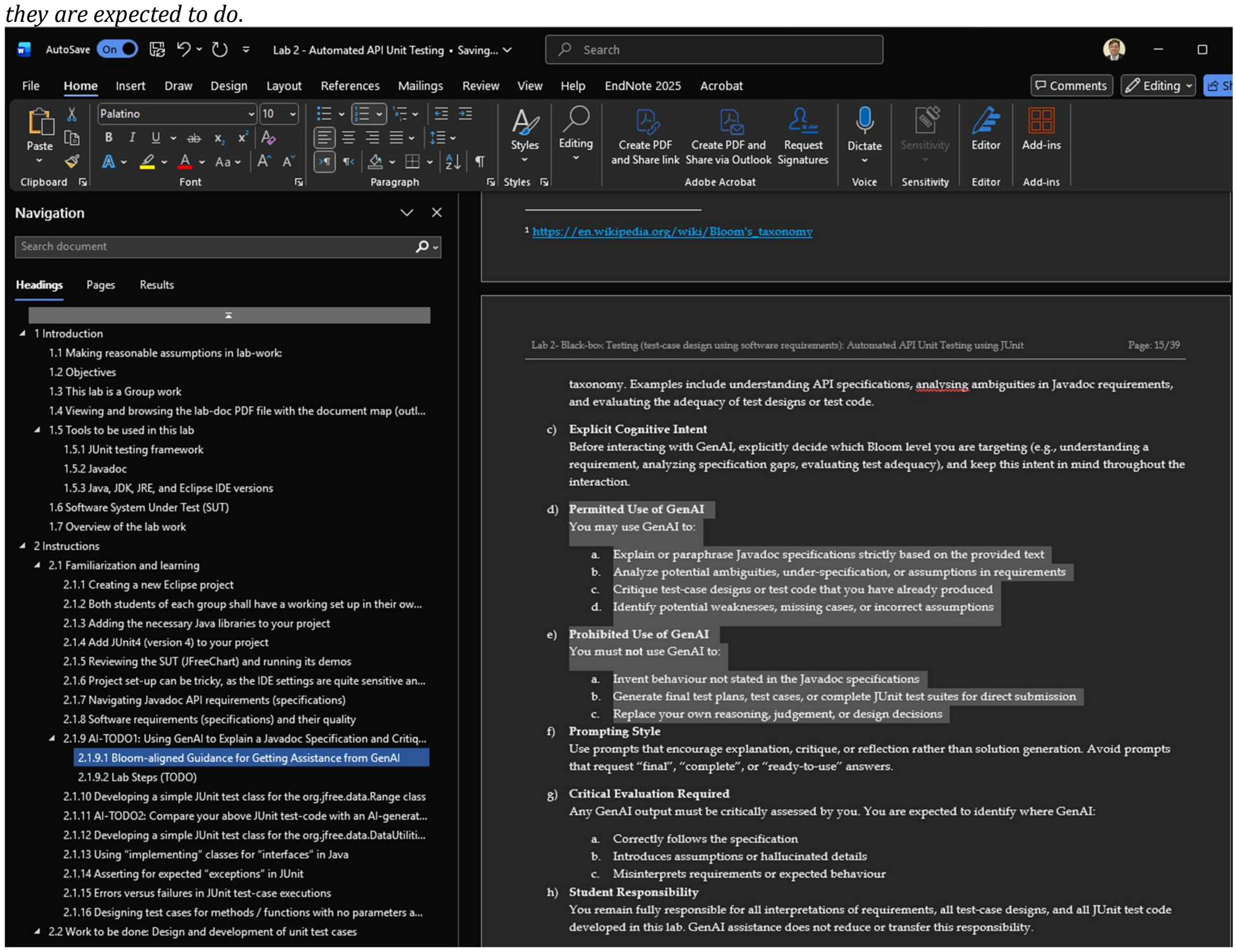

Figure 8-An example of including explicit Bloom-aligned GenAI-usage guidance in laboratory instruction documents. This example is from "*Lab 2 - Automated API Unit Testing*" lab document

In summary, the above three laboratory examples demonstrate how Bloom-aligned GenAI guidance can be operationalized consistently across different types of software testing activities. While the technical focus of the labs varies—from manual and exploratory testing, to test-driven development, to automated API unit testing—the role of GenAI remains deliberately stable. In all cases, GenAI is positioned as a *tutor or mentor* that supports understanding, analysis, and evaluation, rather than as a generator of final solutions. By making the intended cognitive level explicit, constraining permitted and prohibited uses, and requiring students to retain responsibility for all design and judgement-intensive decisions, the lab designs illustrate how GenAI can be integrated in a way that supplements learning across contexts without displacing essential cognitive work. This consistency across labs provides a concrete instantiation of the Bloom-aligned framework and prepares the ground for examining how students perceive and use GenAI under such structured guidance.

### 3.6 Context #2 Implementation: Integrating Bloom Levels and GenAI into AzTU's Revised Course Syllabuses

In the second context of this study, the proposed integration of GenAI is operationalized at the course-syllabus design level, rather than at the level of individual learning activities alone. Building on the revised version of Bloom's taxonomy, GenAI is deliberately positioned as a supporting cognitive tool that can enhance higher-order learning—particularly at the analyze and evaluate levels—while preserving the central role of student reasoning, problem-solving, and reflection.

To demonstrate how this approach can be embedded structurally into formal curricula, we illustrate it through concrete syllabus-level adaptations. Each example shows how GenAI can be explicitly referenced as a learning aid, with clearly defined boundaries, pedagogical intent, and alignment with intended learning outcomes. Importantly, GenAI is not introduced as a shortcut for completing coursework, but rather as a mechanism to help students interrogate their own solutions, compare alternative approaches, and critically assess design or implementation decisions.

As an initial illustration, we show in Figure 9 a screenshot from the syllabus document of the newly-designed Advanced Programming course at AzTU. In this course, students are expected to engage with complex concepts such as abstraction, object-oriented design, architectural trade-offs, and non-trivial implementation decisions. In the revised syllabus for this course, GenAI is incorporated in a controlled and transparent manner to support analytical reasoning and evaluative judgement. For example, students may be encouraged to first develop and justify their own designs or code artefacts, and only then use GenAI to analyze alternative implementations, identify potential weaknesses, or evaluate trade-offs against given criteria. This sequencing ensures that cognitive effort is retained at the appropriate Bloom levels, while GenAI serves as a reflective and comparative aid rather than a substitute for learning.

The second course-level example within Context #2 is drawn from a Data Structures and Algorithms course, where the integration of GenAI is explicitly aligned with higher-order cognitive levels of the revised Bloom's taxonomy. We show a few screenshots from the actual syllabus file in Figure 10. In this syllabus, GenAI is introduced as a structured learning aid to support *analysis* and *evaluation* of algorithmic solutions, rather than as a means of generating answers.

Students are required to first design and implement their own data structures or algorithms and conduct an initial complexity analysis independently. Only after this stage are they encouraged to use GenAI to analyze algorithm behavior, identify potential inefficiencies or edge cases, and compare alternative solutions under different constraints. This sequencing ensures that core algorithmic reasoning—such as Big-O analysis, correctness arguments, and trade-off assessment—remains firmly grounded in the student's own work.

From a pedagogical perspective, this approach offers several benefits. First, it reinforces deep conceptual understanding by prompting students to critically interrogate both their own solutions and AI-generated analyses, rather than accepting automated feedback at face value. Second, it strengthens evaluative judgement, as students must justify why one algorithmic approach is more suitable than another based on formal criteria such as time complexity, space usage, scalability, and problem context. Third, by embedding AI usage explicitly within the syllabus and linking it to Bloom's higher cognitive levels, the course provides clear guidance on responsible and purposeful AI use, reducing ambiguity for both students and instructors.

| English translation of the syllabus, done for the purpose of this paper | 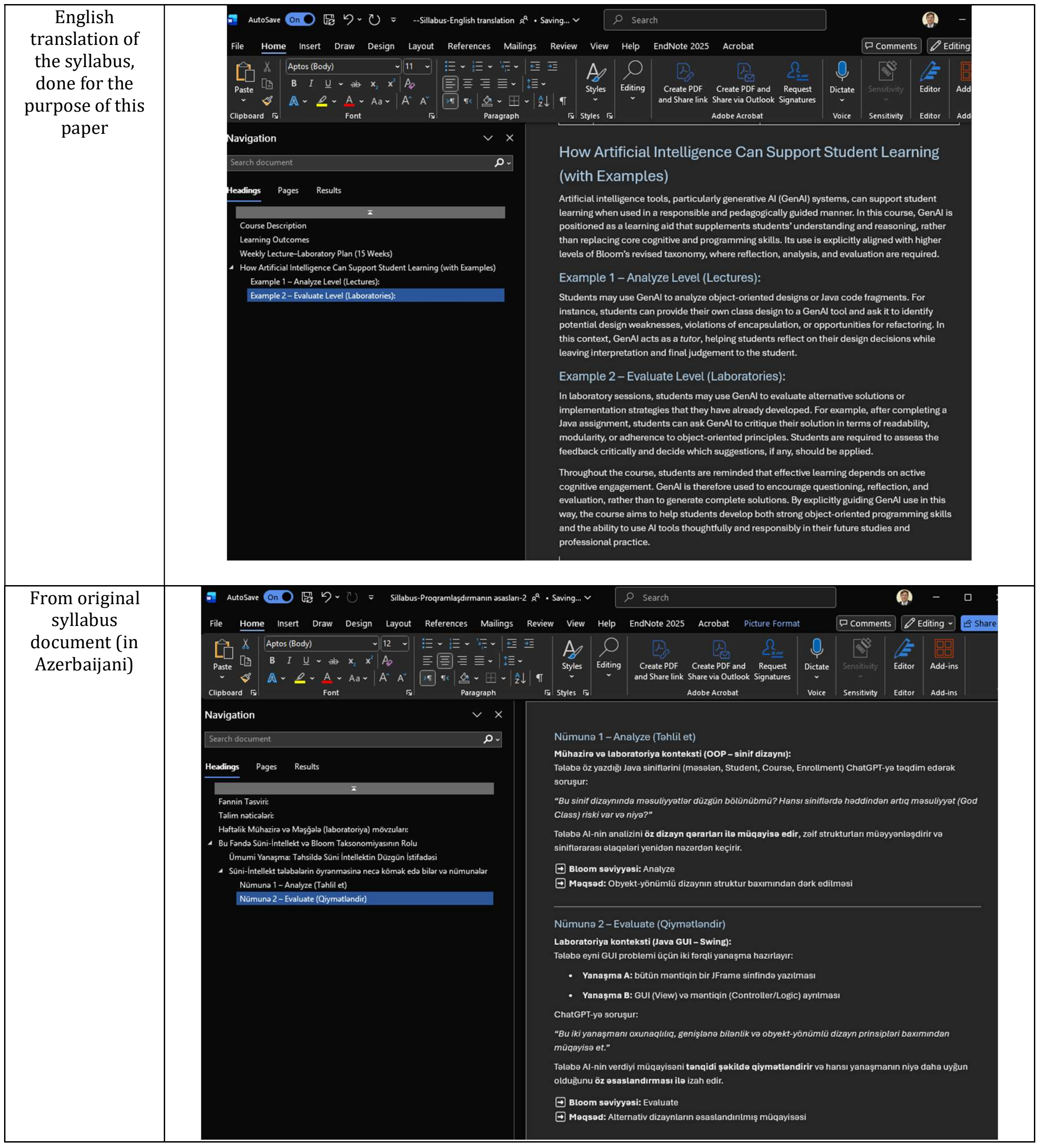 |
| --- | --- |
| From original syllabus document (in Azerbaijani) | |

Figure 9-An example of including explicit Bloom-aligned GenAI-usage guidance in syllabus of AzTU course: Advanced Programming

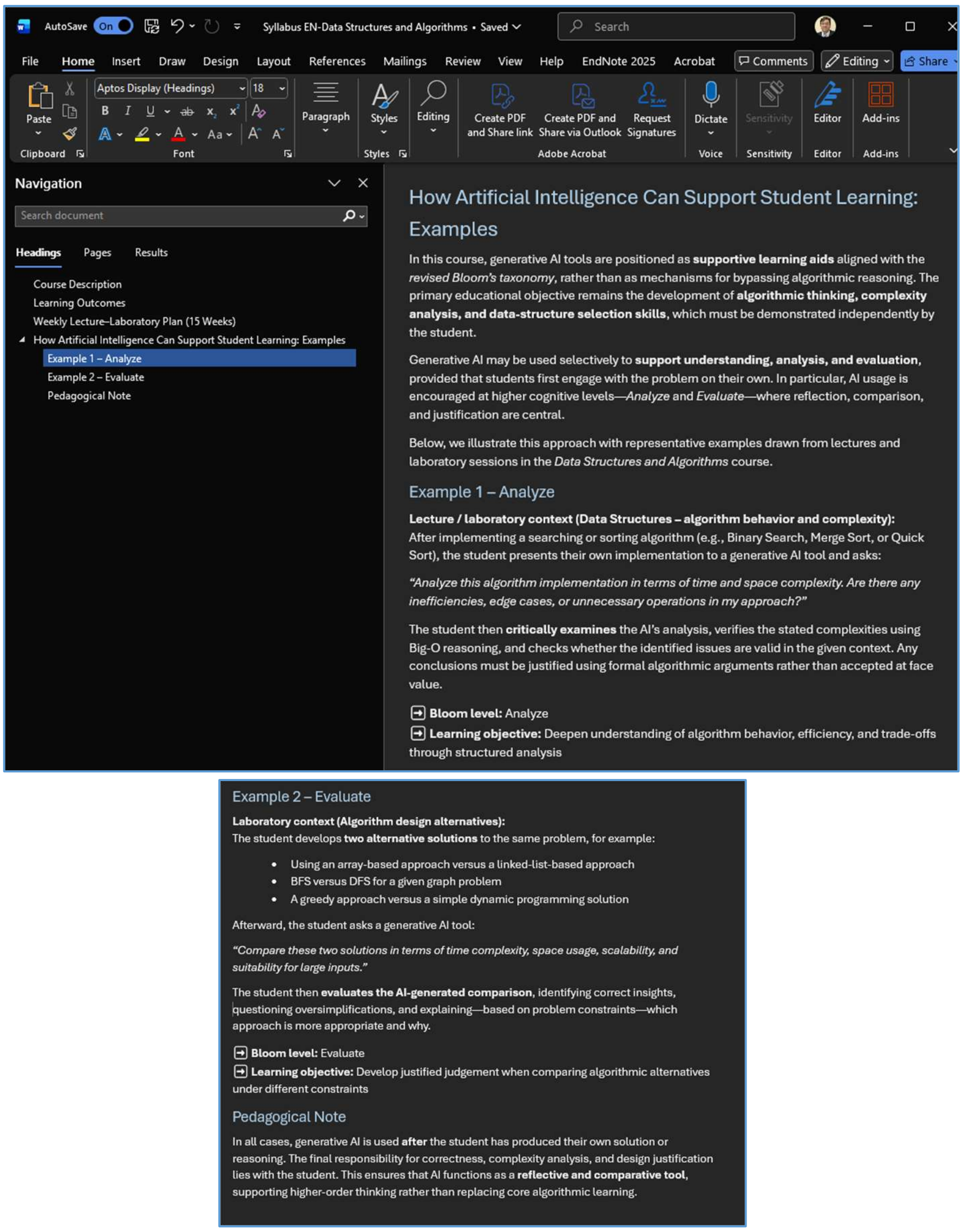

How Artificial Intelligence Can Support Student Learning: Examples

In this course, generative AI tools are positioned as **supportive learning aids** aligned with the *revised Bloom's taxonomy*, rather than as mechanisms for bypassing algorithmic reasoning. The primary educational objective remains the development of **algorithmic thinking, complexity analysis, and data-structure selection skills**, which must be demonstrated independently by the student.

Generative AI may be used selectively to **support understanding, analysis, and evaluation**, provided that students first engage with the problem on their own. In particular, AI usage is encouraged at higher cognitive levels—*Analyze* and *Evaluate*—where reflection, comparison, and justification are central.

Below, we illustrate this approach with representative examples drawn from lectures and laboratory sessions in the *Data Structures and Algorithms* course.

Example 1 – Analyze

**Lecture / laboratory context (Data Structures – algorithm behavior and complexity):**
After implementing a searching or sorting algorithm (e.g., Binary Search, Merge Sort, or Quick Sort), the student presents their own implementation to a generative AI tool and asks:

*"Analyze this algorithm implementation in terms of time and space complexity. Are there any inefficiencies, edge cases, or unnecessary operations in my approach?"*

The student then **critically examines** the AI's analysis, verifies the stated complexities using Big-O reasoning, and checks whether the identified issues are valid in the given context. Any conclusions must be justified using formal algorithmic arguments rather than accepted at face value.

**Bloom level:** Analyze
**Learning objective:** Deepen understanding of algorithm behavior, efficiency, and trade-offs through structured analysis

Example 2 – Evaluate

**Laboratory context (Algorithm design alternatives):**
The student develops **two alternative solutions** to the same problem, for example:

- Using an array-based approach versus a linked-list-based approach
- BFS versus DFS for a given graph problem
- A greedy approach versus a simple dynamic programming solution

Afterward, the student asks a generative AI tool:

*"Compare these two solutions in terms of time complexity, space usage, scalability, and suitability for large inputs."*

The student then **evaluates the AI-generated comparison**, identifying correct insights, questioning oversimplifications, and explaining—based on problem constraints—which approach is more appropriate and why.

**Bloom level:** Evaluate
**Learning objective:** Develop justified judgement when comparing algorithmic alternatives under different constraints

Pedagogical Note

In all cases, generative AI is used **after** the student has produced their own solution or reasoning. The final responsibility for correctness, complexity analysis, and design justification lies with the student. This ensures that AI functions as a **reflective and comparative tool**, supporting higher-order thinking rather than replacing core algorithmic learning.

Figure 10-An example of including explicit Bloom-aligned GenAI-usage guidance in syllabus of AzTU course: Data Structures and Algorithms

### 3.7 Cross-Cutting Design Principles for Bloom-Aligned GenAI Guidance

The implementations described in Sections 3.4 and 3.5 illustrate how Bloom-aligned GenAI guidance was operationalized in two distinct educational contexts, spanning individual course activities and syllabus-level curriculum design. Stepping back from those two concrete instances, this section distils a set of cross-cutting design principles that governed all implementations, independent of course topic, institutional setting, or instructional format. These principles articulate the intended pedagogical role of GenAI, the mechanisms through which Bloom-level alignment was made explicit in teaching practice, and the ways in which responsible GenAI use was framed and discussed with students. By abstracting these principles from the specific cases, we aim to clarify the transferable elements of the approach and provide instructors with a reusable design lens for integrating GenAI into software engineering and computer science education in a learning-oriented and cognitively intentional manner.

#### *3.7.1 GenAI-as-Tutor and Bloom-Level Intentionality*

A central design principle underlying all implementations in this study is the explicit positioning of generative AI as a tutor or mentor, rather than as a solution generator. GenAI was deliberately framed as a cognitive support that prompts reflection, critique, and comparison, while preserving the student's responsibility for core problem-solving, judgment, and decision-making. This principle was applied consistently across courses, learning activities, and instructional artifacts to prevent GenAI from displacing essential cognitive work, particularly in tasks that require analytical reasoning, evaluative judgment, or creative synthesis.

Closely coupled with this tutor-oriented framing is the principle of Bloom-level intentionality. Rather than allowing unrestricted or uniform use of GenAI across all learning tasks, GenAI interactions were explicitly aligned with the intended cognitive level of each activity as defined by Bloom's taxonomy. For lower-level cognitive tasks (e.g., remember and understand), GenAI use was limited or discouraged to avoid undermining foundational knowledge acquisition. In contrast, for higher-level tasks—particularly analyze and evaluate—GenAI was encouraged as a reflective aid to help students inspect assumptions, identify weaknesses, compare alternatives, and justify decisions. This alignment ensured that GenAI use supported, rather than short-circuited, the cognitive demands of the task.

#### *3.7.2 Mechanisms for Embedding Bloom-Aligned GenAI Guidance into Teaching Practice*

To ensure that Bloom-aligned GenAI guidance was consistently enacted rather than implicitly assumed, the design principles were embedded directly into everyday teaching artifacts and practices. This included explicit references to Bloom levels and permitted GenAI roles within lecture slides, laboratory instructions, syllabus documents, and assignment briefs. By making expectations visible at the point of task execution, students were guided toward intentional and level-appropriate GenAI use without relying on informal norms or ad hoc instructor clarification.

In addition, instructors reinforced this guidance through repeated verbal framing during lectures and labs, emphasizing GenAI's role as a reflective support rather than a source of final answers.

#### *3.7.3 Instructor–Student Meta-Discussions on Responsible GenAI Use*

A further cross-cutting principle of the implementation was the deliberate inclusion of explicit meta-discussion between instructors and students about the role, value, and risks of GenAI in learning. Rather than treating GenAI use as implicit or purely regulated through policy, instructors openly discussed why and when GenAI should be used, how it relates to Bloom's cognitive levels, and where its use may undermine learning. Instructors consistently framed GenAI as a learning supplement, not a replacement for thinking, emphasizing messages such as: "GenAI can help you think better about your solution, but it cannot replace the thinking you are expected to do," and "If GenAI does the core reasoning for you, then the learning objective has already been missed."

Students were encouraged to reflect on their own GenAI usage practices and to articulate perceived benefits and risks. Typical student comments included observations such as: "Using GenAI after I finished my work helped me see gaps in my reasoning," and "When I asked GenAI too early, I realized I was skipping the hard thinking part." Others explicitly linked the discussion to Bloom's taxonomy, noting that "GenAI made more sense for analysis and evaluation than for writing the initial solution." These exchanges helped normalize transparent GenAI use, reduce ambiguity about acceptable practices, and promote intentional, level-appropriate engagement. Overall, the instructor–student meta-discussion functioned as a pedagogical mechanism for cultivating shared norms around responsible GenAI use and reinforcing student accountability across learning activities.

### 3.8 Summary of the Implementation Approach

In summary, we implemented a Bloom-aligned framework for GenAI-supported learning across two educational contexts—an individual course setting and a faculty-wide course update—using a consistent and intentional design strategy. Rather than introducing GenAI as a generic productivity tool, the approach explicitly framed GenAI as a tutor or mentor whose role varies according to the cognitive demands of learning tasks. Bloom's taxonomy served as the organizing structure for determining when, how, and for what purpose GenAI could supplement student learning.

Across all implementations, three core elements characterized the approach: (1) explicit alignment between GenAI use and Bloom's cognitive levels, with particular emphasis on higher-order thinking; (2) clear embedding of this guidance into teaching artifacts such as lectures, labs, syllabuses, and assignments; and (3) open instructor–student meta-discussion to establish shared expectations and responsible use norms. Together, these elements ensured that GenAI use was transparent, intentional, and learning-oriented, while preserving student agency and accountability. This implementation framework provides the foundation for the empirical assessment presented in Section 4, where we examine how Bloom-aligned GenAI guidance influenced student learning support and usage practices.

## 4 EMPIRICAL ASSESSMENT OF THE APPROACH

This section presents the empirical assessment of the proposed Bloom-aligned framework for GenAI-supported learning. The goal of this assessment is to examine how the approach was experienced and used in practice, rather than to establish causal learning effects. Specifically, we investigate: (1) the cognitive levels at which students perceived GenAI to support their learning;, (2) how explicit Bloom-level guidance influenced students' GenAI usage practices; and (3) the perceived benefits and drawbacks of the approach from both student and instructor perspectives. To this end, we adopt a qualitative research design that combines questionnaire data and selected learning artifacts. The following subsections describe the research questions, data collection instruments, analysis procedures, and ethical considerations, followed by the empirical results.

### 4.1 Goal and Research Questions

The goal of this empirical assessment is to examine how the proposed Bloom-aligned, explicitly guided GenAI approach was experienced and used in the two educational settings (QUB in the UK and AzTU in Azerbaijan). Rather than attempting to establish that the approach caused improvements in student learning, the assessment focuses on understanding where and how GenAI-supported learning, how explicit Bloom-level guidance shaped students' GenAI usage practices, and what benefits and drawbacks were perceived by students and instructors.

Establishing objective or causal learning gains attributable solely to GenAI-supported interventions is inherently challenging in real-world educational contexts. Student learning outcomes are influenced by a wide range of confounding factors, including prior knowledge, individual motivation, learning strategies, instructor style, assessment design, peer interaction, workload from other courses, and concurrent use of external learning resources. In addition, GenAI tools themselves evolve rapidly, students' familiarity with these tools varies considerably, and GenAI usage often occurs alongside traditional learning activities rather than in isolation. In principle, one possible way to assess causal learning effects would be to conduct a controlled experiment by dividing students into a control group and an experimental group with carefully matched backgrounds, prior expertise, and learning conditions. However, implementing such tightly controlled experimental designs in authentic academic settings—across real courses, real semesters, and real assessment constraints—is extremely difficult and often impractical. As a result, attributing observed performance changes uniquely or reliably to the proposed approach without introducing substantial validity threats is rarely feasible in practice.

Given these constraints, this study adopts a practice-oriented and experience-focused evaluation strategy. The emphasis is on examining students' perceived learning support across Bloom's cognitive levels, observable changes in GenAI usage practices when guidance is made explicit, and reflective accounts of the approach's benefits and limitations. This perspective aligns with the paper's pedagogical objective: to understand how GenAI can supplement learning when intentionally framed, rather than to claim definitive learning gains under controlled conditions.

Accordingly, the study is guided by the following research questions:

- **RQ1** (**Bloom-Level Learning Support)**: At which Bloom cognitive levels did students report that GenAI supported their learning, and in what ways did that support differ across levels?
- **RQ2** (**GenAI Usage Practices**): How does explicit Bloom-level guidance influence students' GenAI usage practices during learning activities?
- **RQ3 (Perceived Benefits and Drawbacks):** What benefits and limitations do students and instructors perceive in using a Bloom-aligned, explicitly guided GenAI approach to support learning in software engineering and computer science?

### 4.2 Research Method

This study adopts a qualitative research design to address the research questions defined in Section 4.1, by analyzing the qualitative evidence collected from the two educational settings (contexts), as discussed in Section 3.1. The objective is to assess how the proposed Bloom-aligned GenAI approach was experienced and used in practice, rather than to establish causal learning effects.

Table 5 maps each research question to its corresponding data sources and analysis approach, making explicit the alignment between questions, evidence, and analysis and supporting methodological transparency. For RQ1, data are drawn primarily from student questionnaires with short reflections and selected learning artifacts.

For RQ2, evidence consists of student self-reports, reflective responses, and indicators embedded in submitted coursework artifacts. These data are analyzed using descriptive and comparative analysis of usage practices, together with thematic analysis of reflections. For RQ3, qualitative data from open-ended student responses and instructor reflections are analyzed thematically to identify perceived benefits and drawbacks and to compare perspectives across stakeholders.

Table 5- Mapping of Research Questions to Data Sources and Analysis Approach

| Research Question | Data Sources | Analysis Approach |
|---|---|---|
| **RQ1 – Bloom-Level Learning Support:** At which Bloom cognitive levels did students report that GenAI supported their learning, and how did this support differ across levels? | Student questionnaires with open-ended student reflections, and selected learning artifacts for contextual interpretation | Qualitative coding of explanatory comments |
| **RQ2 – GenAI Usage Practices:** How does explicit Bloom-level guidance influence students' GenAI usage practices during learning activities? | Student self-reports on GenAI usage practices, short reflective responses, and evidence from submitted coursework artifacts | Descriptive and comparative analysis of reported usage patterns, thematic analysis of reflections, and triangulation between self-reported practices and artifact indicators |
| **RQ3 – Perceived Benefits and Drawbacks:** What benefits and limitations do students and instructors perceive in using a Bloom-aligned, explicitly guided GenAI approach? | Open-ended student questionnaire responses and instructor reflections collected during and after course delivery | Qualitative thematic analysis to identify perceived benefits and drawbacks, followed by cross-stakeholder comparison between student and instructor perspectives |

#### *4.2.1 Designing the Anonymous Questionnaires*

To address the research questions, we designed a set of fully qualitative, anonymous questionnaires to capture reflective accounts of students' and instructors' experiences with the Bloom-aligned GenAI approach. We designed the questionnaires to have open-ended, free-text questions only, intentionally avoiding Likert-scale items in order to elicit rich descriptions of learning support, usage practices, and perceived benefits and drawbacks.

The questionnaires were aligned explicitly with the research questions. To address RQ1, a student questionnaire prompted participants to describe how GenAI supported (or did not support) their learning at each Bloom cognitive level. To address RQ2, a separate student questionnaire focused on how explicit Bloom-level guidance influenced GenAI usage practices, including when and how GenAI was used. To address RQ3, qualitative reflection questionnaires were used to collect perceived benefits and limitations from both student and instructor perspectives.

A short qualitative background questionnaire captured students' prior experience with GenAI tools to support contextual interpretation. Table 6 summarizes the mapping between questionnaires, research questions, and empirical contexts.

Table 6- Mapping of Questionnaires to Research Questions and Empirical Contexts

| Questionnaire | Target Research Question(s) | Empirical Context |
|---|---|---|
| 1-Bloom-Level Learning Support Questionnaire (Students) | RQ1 | QUB – Software Testing course |
| 2-GenAI Usage Practices Questionnaire (Students) | RQ2 | QUB – Software Testing course |
| 3-Perceived Benefits and Drawbacks Questionnaire (Students) | RQ3 | QUB – Software Testing course |
| 4-Instructor Reflection Questionnaire | RQ3 | AzTU – redesigned course syllabuses |

| Student Background and GenAI Familiarity Questionnaire | Contextual (RQ1–RQ3) | QUB – Software Testing course |
|---|---|---|

#### *4.2.2 Implementations of the Anonymous Questionnaires*

All questionnaires were administered anonymously and were hosted using Google Forms. No personally identifiable information was collected, and responses could not be linked to individual participants. Anonymity was explicitly emphasized at the time of data collection to encourage honest and reflective responses, particularly with respect to GenAI usage practices, perceived learning support, and reported challenges.

The student questionnaires were administered after students had substantial exposure to the Bloom-aligned GenAI guidance during the course, ensuring that responses reflected actual learning experiences rather than expectations or assumptions. Questionnaires were distributed by email to the entire class, and it was clearly communicated that participation was entirely voluntary and anonymous. Students were given sufficient time to complete the questionnaires outside scheduled teaching activities, reducing time pressure and potential response bias.

Instructor reflection questionnaires were administered separately after the redesigned syllabuses had been used in teaching. These reflections were collected independently of student data and focused on instructional experiences, feasibility, and observed challenges associated with implementing the approach.

A total of 42 students completed the anonymous student questionnaires in the Software Testing course at Queen's University Belfast, corresponding to a participation rate of approximately 27% (42 out of 153 students). For the instructor reflection questionnaire, instructors of 11 out of the 18 newly redesigned courses at Azerbaijan Technical University provided responses, representing substantial instructor engagement with the redesigned syllabuses.

#### *4.2.3 Ethical Implications and Considerations*

Ethical considerations were carefully addressed throughout the design and execution of the study. Participation in all questionnaires was entirely voluntary and was preceded by informed consent. Participants were clearly informed about the purpose of the study, the fully anonymous nature of data collection, and their right to withdraw at any time without any negative consequences. No personally identifiable information was collected, and questionnaire responses were not linked to individual students or instructors in any form. In addition, no incentives were offered for participation, and involvement in the study had no impact on course assessment, grading, academic standing, or professional evaluation.

Given the anonymous design of the questionnaires, the voluntary nature of participation, and the absence of sensitive or personally identifiable data, formal institutional ethics approval was not required under the applicable institutional guidelines for educational research at the participating institutions. To ensure due diligence, the study design and data collection procedures were reviewed against the research ethics guidelines of all three institutions represented by the co-authors, and none required a formal ethics application or approval for this type of educational research.

The primary ethical emphasis of the study was placed on transparency, informed participation, and the protection of participant autonomy. This was particularly important given the reflective nature of the questions, which invited participants to comment on their own GenAI usage practices and learning behaviors. Care was taken to frame questions in a non-judgmental manner and to avoid any language that could be perceived as evaluative or punitive. Overall, these measures ensured that the study adhered to established ethical principles for educational research while minimizing risk to participants and preserving trust between instructors and learners.

#### *4.2.4 Data Analysis Approach*

The data analysis followed a qualitative, research-question–driven approach, consistent with the qualitative research design outlined at the beginning of Section 4. Analysis strategies were selected to align directly with each research question, as summarized in Table 5, and combined evidence from open-ended questionnaire responses and selected learning artifacts.

For RQ1, students' qualitative responses describing how GenAI supported (or did not support) their learning across Bloom's cognitive levels were analyzed using thematic analysis, with Bloom's taxonomy serving as an analytic lens rather than a measurement framework. In addition to questionnaire responses, selected learning artifacts (e.g., reflective lab submissions) were used to contextualize and corroborate students' descriptions of GenAI-supported learning at higher cognitive levels such as Analyze and Evaluate.

For RQ2, qualitative responses related to GenAI usage practices were thematically analyzed to identify patterns in when, how, and why students used GenAI under explicit Bloom-level guidance. This analysis was complemented by examination of selected learning artifacts to triangulate reported practices with observable indicators of reflective or evaluative GenAI use.

For RQ3, qualitative reflections from students and instructors were analyzed using inductive thematic analysis to identify perceived benefits and drawbacks of the approach. Themes were iteratively refined and compared across stakeholder groups to capture convergent and divergent perspectives on pedagogical value and limitations.

To support the qualitative analysis process, a GenAI tool (OpenAI's ChatGPT) was used as an analytical aid for initial code suggestion, theme clustering, and summarization. All AI-assisted outputs were reviewed, refined, and validated by the researchers, who retained full responsibility for coding decisions, interpretation, and reporting. This use of GenAI was intended to support analytic efficiency and consistency while preserving human judgment.

### 4.3 Results

We present and analyze the results of each RQ next.

#### *4.3.1 RQ1-Extent of GenAI Support Across Bloom Levels*

RQ1examined how students perceived GenAI to support their learning across Bloom's cognitive levels, based on qualitative responses from 42 students who completed the anonymous questionnaires in the Software Testing course. The analysis focused on identifying recurring themes, contrasts, and explanatory patterns in students' descriptions of GenAI-supported learning.

At the lower cognitive levels of *Remember* and *Understand*, students consistently reported limited or conditional usefulness of GenAI. Many students emphasized that early or excessive reliance on GenAI interfered with the formation of foundational understanding. One student reflected: *"Using GenAI too early made it harder to really understand the basics. When I asked it before trying myself, I felt like I was reading answers without actually processing why things worked."*

This sentiment was echoed by other students, who described a sense of *illusory understanding* when GenAI was used prematurely. For example, another student noted: *"It gave explanations, but I didn't feel confident I actually understood them unless I struggled with the material first."*

These reflections suggest that students perceived GenAI use at lower Bloom levels as potentially counterproductive, unless carefully sequenced after initial independent learning. This aligns with the instructional guidance that discouraged GenAI use for basic knowledge acquisition and emphasizes the importance of "productive struggle" [29] at early learning stages.

At the *Apply* level, students reported a more nuanced role for GenAI. Rather than generating solutions, GenAI was described as a confirmation and clarification tool, helping students assess whether they were applying known techniques appropriately. One student explained: *"After applying the testing technique myself, I used GenAI to see if my interpretation made sense. It didn't replace my work, but it reassured me that I was on the right track."*

Such comments indicate that GenAI was perceived as supportive when used *after* students had engaged with the task independently, reinforcing correct application rather than bypassing learning.

The strongest and most consistently positive perceptions of GenAI support were associated with the higher-order cognitive levels, particularly *Analyze* and *Evaluate*. Students frequently described GenAI as enabling deeper reflection on their own work by exposing blind spots, assumptions, and alternative perspectives. One student elaborated: *"Once I had my solution, GenAI helped me analyze it more critically. It pointed out assumptions I didn't realize I was making, which made me rethink parts of my answer".* Another student emphasized its evaluative role: *"I didn't use it to write my solution, but to compare my approach with other possible ones and judge whether mine was strong or weak."*

These accounts suggest that students experienced GenAI as a cognitive mirror, supporting metacognitive awareness and evaluative judgment rather than content production.

At the *Create* level, perceptions were more mixed. Some students valued GenAI for ideation and inspiration, particularly when exploring alternative test scenarios or edge cases, while others expressed caution about over-reliance. One student summarized this tension clearly: *"It was useful for brainstorming ideas, but I didn't feel comfortable letting it create anything final. I still wanted the creative part to be mine."*

Some students also reported feeling tempted to use GenAI for direct answer or artifact generation, particularly when tasks were demanding or time-pressured. Several noted that resisting this temptation was not always easy, as GenAI could quickly produce fluent outputs. As one student reflected, *"It was tempting to let the AI generate the answer when I was stuck."* However, some students described deliberately avoiding GenAI during initial problem solving and using it only after engaging analytically with the task. These reflections suggest that, despite the temptation to delegate cognitive work, some students exercised self-regulation and aligned their GenAI use with the intended pedagogical framing.

#### *4.3.2 RQ2-Effects of Explicit Bloom-Level Guidance on Students GenAI Usage Practices*

RQ2 examined how explicit Bloom-level guidance influenced students' GenAI usage practices, based on qualitative responses from 42 students in the Software Testing course. The analysis focused on identifying changes in when, how, and why students chose to interact with GenAI during learning activities, as well as instances in which students deliberately avoided its use.

A dominant theme across responses was a shift in the timing of GenAI use. Many students reported that, prior to receiving Bloom-level guidance, they tended to consult GenAI early in the task-solving process. After explicit guidance was introduced, students described postponing GenAI interaction until after completing an initial solution or forming their own reasoning. One student explained: *"Before the guidance, I would often ask GenAI straight away. Afterward, I tried to solve the problem first and only used it once I had something to check."*

This shift reflects a move from exploratory or convenience-driven usage toward more intentional, reflective engagement. Students also reported changes in how they used GenAI. Rather than requesting full solutions, many described using GenAI to critique, validate, or stress-test their own work. Typical comments included: *"I stopped asking it to generate answers and instead asked it to analyze my solution";* and *"I used GenAI more like a reviewer, not a solver."*

These responses suggest that Bloom-level guidance helped students reconceptualize GenAI as a support for analysis and evaluation, rather than as a shortcut to task completion.

Another recurring theme was intentional non-use. Several students described situations in which they consciously decided not to use GenAI, particularly for tasks targeting lower cognitive levels or for initial reasoning. One student noted: *"Once I understood that some tasks were meant to test my own thinking, I avoided using GenAI, even though it was available."*

This indicates that explicit guidance did not simply shape GenAI usage, but also legitimized decisions not to use GenAI as part of responsible learning practice.

#### *4.3.3 RQ3- Perceived Benefits and Drawbacks of the Approach*

This section reports the perceived benefits and drawbacks of the Bloom-aligned, explicitly guided GenAI approach, based on qualitative reflections from students in the Software Testing course and instructors involved in the redesigned courses at AzTU. The findings are organized by stakeholder group and valence to clearly distinguish perspectives and experiences.

Perceived Benefits for Students

A central benefit reported by students was increased clarity and intentionality in using GenAI for learning. Many students described the Bloom-level framing as providing a concrete rationale for when GenAI use was pedagogically meaningful, rather than relying on vague rules or prohibitions. One student reflected, "The Bloom levels helped me understand why GenAI made sense for some tasks but not others."

Students also reported enhanced self-regulation and metacognitive awareness. Several described becoming more reflective about their learning process and more deliberate in deciding whether GenAI use would genuinely support understanding. As one student noted, "I started thinking more about whether GenAI was actually helping me learn or just saving time." These reflections suggest that the approach supported students in developing a more thoughtful and learning-oriented relationship with GenAI.

Perceived Drawbacks for Students

Despite these benefits, students also identified several challenges. A recurring concern was the initial learning curve associated with understanding and applying Bloom-level guidance to their own GenAI usage. Especially early in the course, some students found it cognitively demanding to reflect on Bloom levels in parallel with solving technical tasks. One student commented, "At first, it felt like extra effort to think about Bloom levels on top of the task itself."

Another commonly mentioned drawback was the ongoing temptation to use GenAI for initial solution or artifact generation, even when students recognized that doing so might undermine learning. This temptation was sometimes amplified during the early learning curve, when students were still internalizing the intent of the guidance. As one student acknowledged, "Even knowing the rules, it was sometimes hard not to let GenAI do the first draft." These accounts indicate that while the guidance shaped behavior over time, it did not immediately eliminate tensions between convenience and learning.

Perceived Benefits for Instructors

From the instructors' perspective, a key benefit of the approach was its ability to reframe GenAI discussions in pedagogical terms. Rather than focusing on detection or enforcement, instructors reported that Bloom's taxonomy provided a shared language for discussing appropriate GenAI use. One instructor observed, "It shifted the conversation from policing AI use to talking about learning objectives."

Instructors also noted improvements in the quality of student reasoning and reflection, particularly in tasks targeting analysis and evaluation. Several instructors reported that students were more explicit about assumptions and justification when GenAI was framed as a tool for critique rather than generation. This was perceived as a positive shift toward deeper engagement with course material.

Perceived Drawbacks for Instructors

Instructors also identified notable limitations. A frequently cited drawback was the learning curve and additional preparation effort required to design, communicate, and consistently reinforce Bloom-aligned GenAI guidance across teaching materials. One instructor explained, "Designing tasks and explanations with Bloom levels in mind definitely takes more time."

Instructors further emphasized that the approach does not fully prevent inappropriate GenAI use and requires sustained reinforcement, particularly while both instructors and students are still progressing along the learning curve. Some noted that without ongoing discussion and reminders, students could revert to convenience-driven usage. These reflections highlight that the approach is not a one-time intervention, but an evolving pedagogical practice that matures over time.

# 5 DISCUSSIONS, KEY INSIGHTS, IMPLICATIONS, AND LIMITATIONS

## 5.1 Discussions and Insights

The findings highlight that the perceived value of GenAI is highly dependent on cognitive level and timing of use, rather than on access alone.

Students consistently reported limited or negative learning value when GenAI was used at lower Bloom levels, particularly during initial knowledge acquisition. In contrast, GenAI was perceived as most useful at higher cognitive levels—especially *Analyze* and *Evaluate*—where it supported critique, comparison, and reflective reasoning. This reinforces Bloom's taxonomy as a practical framework not only for learning objectives, but also for structuring responsible GenAI use.

A key insight is the importance of explicit pedagogical framing. Bloom-level guidance helped students rethink *when* and *how* to engage with GenAI, encouraging delayed use, reflective prompting, and intentional avoidance during early problem solving. Rather than enforcing compliance, the guidance shaped students' intent and supported self-regulation.

The findings also reveal a clear learning curve for both students and instructors. Students described cognitive effort and temptation, particularly early on, while instructors reported increased preparation and communication effort. However, these challenges tended to lessen as expectations became clearer and practices stabilized.

From an instructional perspective, Bloom-aligned guidance shifted GenAI-related discussions from policing toward pedagogy, providing a shared language for linking GenAI use to learning goals. Overall, the results suggest that GenAI's educational value lies not in unrestricted use or prohibition, but in intentional alignment between cognitive goals, guidance, and learner self-regulation.

Key take-away messages are:

- GenAI is perceived as most valuable for higher-order cognitive activities, not foundational learning.
- Explicit Bloom-level guidance meaningfully influences students' GenAI usage practices.
- Both students and instructors experience a learning curve, requiring sustained guidance.
- Framing GenAI as a tutor rather than a solution generator supports deeper learning.
- Bloom's taxonomy offers a scalable framework for pedagogy-driven GenAI integration in SE/CS education.

## 5.2 Implications

Building on the empirical results and discussion in the preceding sections, this subsection discusses the implications of the study's findings for key stakeholders involved in software engineering and computer science education.

### *5.2.1 Implications for SE/CS Educators*

For educators in software engineering and computer science, the findings suggest that effective GenAI integration requires instructional design, not policy enforcement alone. Rather than issuing generic rules about allowed or disallowed GenAI use, educators can benefit from explicitly aligning GenAI guidance with cognitive learning goals, using Bloom's taxonomy as a shared reference point. This approach enables instructors to articulate why GenAI is appropriate for certain learning activities and why it should be restricted or delayed for others.

The results also imply that educators should anticipate and plan for a learning curve, both for themselves and for students. Embedding Bloom-aligned GenAI guidance may require additional upfront effort in task design, examples, and classroom discussion. However, this investment can reduce ambiguity, support reflective learning, and shift instructional focus from monitoring misuse to supporting learning processes. Importantly, GenAI guidance should be treated as an ongoing pedagogical practice, reinforced throughout the course rather than introduced as a one-time policy statement.

### *5.2.2 Implications for Education Researchers*

For education researchers, this study highlights the value of examining GenAI use through process-oriented and theory-informed lenses, rather than focusing solely on outcome-based measures such as performance gains. Bloom's taxonomy proved useful not as a measurement instrument, but as an analytic and design framework for understanding how learners interact with GenAI at different cognitive levels.

The findings also suggest that qualitative approaches are particularly well suited for studying GenAI-supported learning, especially when ethical constraints, contextual variability, and rapid technological change make controlled experiments difficult. Researchers may benefit from focusing on learners' reasoning, self-regulation, and decision-making processes, as well as instructors' pedagogical sense-making. More broadly, the study underscores the need for research designs that acknowledge the dynamic and negotiated nature of GenAI use in authentic educational settings.

### 5.2.3 *Implications for Students and Learners*

For students and learners, the findings suggest that GenAI is most beneficial when used as a support for thinking, rather than as a shortcut to completed work. Developing awareness of cognitive goals—such as whether a task aims to build foundational understanding or higher-order reasoning—can help learners make more informed decisions about when and how to use GenAI.

The results also imply that responsible GenAI use involves self-regulation and reflection, particularly in resisting the temptation to delegate initial problem solving to AI tools. While this may introduce short-term effort, learners stand to gain deeper understanding and stronger analytical skills over time. By viewing GenAI as a tutor or critical partner rather than a solution generator, students can better align their GenAI use with long-term learning goals and professional development in SE/CS contexts.

## 5.3 Limitations and Potential Threats to Validity

Given the empirical nature of this research, the findings should be interpreted in light of several limitations and potential threats to validity. Like any empirical study, the investigation reported in Section 4 is subject to constraints related to context, methodology, and data sources. In the following, we discuss these issues using four commonly adopted categories of validity threats [30]—internal validity, construct validity, conclusion validity, and external validity—and outline steps we have taken to mitigate them where possible.

**Internal validity.** Threats to internal validity arise from factors that may influence participants' responses independently of the studied intervention. In this study, students' prior experience with GenAI tools, individual learning strategies, workload pressure, and course-specific characteristics may have affected how they perceived and used GenAI, independent of the Bloom-aligned guidance. While such confounding factors cannot be fully controlled in authentic educational settings, we mitigated their impact by collecting reflective, experience-based responses after sustained exposure to the approach, rather than relying on hypothetical or short-term impressions. Furthermore, the emphasis on qualitative interpretation reduced the risk of attributing causal effects where none could be reliably established.

**Construct validity.** Construct validity concerns whether the study accurately captured the concepts it intended to examine. Concepts such as "GenAI-supported learning," "Bloom-level alignment," and "responsible use" are inherently abstract and context-dependent. To mitigate this threat, the questionnaires were designed using explicit Bloom-level prompts and concrete examples, reducing ambiguity in interpretation. In addition, Bloom's taxonomy was used as an analytic lens rather than a measurement scale, avoiding the risk of oversimplifying complex cognitive processes into numerical indicators.

**Conclusion validity.** Conclusion validity relates to the extent to which the conclusions drawn are reasonable given the data and analysis. As this study employed a qualitative research design with a moderate number of participants, the findings do not support statistical generalization or causal claims about learning effectiveness. To mitigate this threat, we deliberately framed our conclusions in descriptive and interpretive terms, focusing on perceived support, usage practices, and experiences rather than learning gains. The use of thematic analysis, triangulation with selected learning artifacts, and researcher review of AI-assisted coding outputs further strengthened the credibility of the interpretations.

**External validity.** External validity concerns the generalizability of the findings beyond the studied contexts. The student data were collected from a single Software Testing course at one institution, and instructor reflections were drawn from a specific set of redesigned courses at another institution. As such, the findings may not directly transfer to other disciplines, educational levels, or institutional cultures. However, the study's goal was not broad generalization, but analytic transferability. By grounding the approach in a widely recognized framework (Bloom's taxonomy) and reporting rich contextual detail, we enable educators and researchers to assess the relevance and applicability of the findings to their own settings.

In summary, while these limitations are inherent to empirical research conducted in real educational environments, the study provides credible and transparent insights into how Bloom-aligned GenAI guidance can shape learning practices. Future research can build on these findings through longitudinal studies, mixed-method designs, or controlled interventions where feasible.

## 6 CONCLUSIONS AND FUTURE WORK

This paper presented and empirically examined a Bloom-aligned approach for guiding responsible and learning-oriented use of generative AI (GenAI) in software engineering and computer science education. Rather than treating GenAI as either a prohibited shortcut or a neutral productivity tool, the proposed approach positions GenAI as a cognitively intentional learning aid, whose value depends on when, how, and for what cognitive purposes it is used.

Through qualitative evidence from student questionnaires, instructor reflections, and selected learning artifacts, the study demonstrated that explicit Bloom-level guidance can meaningfully shape learners' GenAI usage practices. Students reported perceiving GenAI as most beneficial for higher-order cognitive activities, such as analysis and evaluation, while also developing greater awareness of when GenAI use may undermine foundational learning. Instructors highlighted the value of Bloom's taxonomy as a shared pedagogical language that enabled a shift from enforcement-oriented discussions toward learning-centered guidance. At the same time, the findings revealed a clear learning curve and implementation effort for both students and instructors, underscoring that responsible GenAI integration is an ongoing pedagogical process rather than a one-time intervention.

Taken together, the results suggest that the educational value of GenAI lies not in unrestricted access or strict prohibition, but in intentional alignment between cognitive goals, instructional guidance, and learner self-regulation. Bloom's taxonomy provides a familiar and theoretically grounded framework through which such alignment can be operationalized in SE/CS education.

Future work can extend this research in several directions. Longitudinal studies could examine how students' GenAI usage practices and self-regulation evolve over time as Bloom-aligned guidance becomes internalized. Mixed-method or quasi-experimental designs may help explore potential relationships between guided GenAI use and learning outcomes, where contextual constraints permit. Future studies could also investigate how the approach scales across different disciplines, educational levels, and institutional cultures, as well as how it interacts with assessment design and academic integrity policies. Finally, further research is needed to explore how instructors can be supported in overcoming the learning curve associated with adopting and sustaining Bloom-aligned GenAI guidance in practice.

## ACKNOWLEDGEMENTS

This work was supported by the institutional research grant #2025-1053/25 entitled "*Integration of AI and Bloom's Taxonomy in Computing Education*" by Azerbaijan Technical University. The authors would like to thank the students and instructors of the courses in which the approach presented in this paper was applied.

## REFERENCES

[1] B. S. Bloom, M. D. Engelhart, E. J. Furst, W. H. Hill, and D. R. Krathwohl, *Taxonomy of educational objectives: The classification of educational goals. Handbook 1: Cognitive domain*. Longman New York, 1956.

[2] D. R. Krathwohl, "A revision of Bloom's taxonomy: An overview," *Theory into practice,* vol. 41, no. 4, pp. 212–218, 2002.

[3] Oregon State University. "Bloom's Taxonomy Revisited – Artificial Intelligence Tools." ecampus.oregonstate.edu/faculty/artificial-intelligence-tools/ blooms-taxonomy-revisited (accessed Jan. 2026).

[4] C. E. Smith *et al.*, "Early adoption of GenAI in computing education: Emergent student use cases and perspectives," in *Proceedings of the on Innovation and Technology in Computer Science Education*, 2024, pp. 3–9.

[5] I. Reihanian, Y. Hou, and Q. Sun, "From Pilots to Practices: A Scoping Review of GenAI-Enabled Personalization in Computer Science Education," *AI,* vol. 7, no. 1, p. 6, 2025.

[6] C. Zastudil, M. Rogalska, C. Kapp, J. Vaughn, and S. MacNeil, "Generative AI in Computing Education: Perspectives of students and instructors," in *IEEE Frontiers in Education Conference*, 2023, pp. 1–9.

[7] J. Roberts and A. Mohamed, "Generative AI in CS Education: Literature Review through a SWOT Lens," in *Proceedings of the Western Canadian Conference on Computing Education*, 2024, pp. 1–6.

[8] E. Dickey, A. Bejarano, and C. Garg, "Innovating Computer Programming Pedagogy: The AI-Lab Framework for Generative AI Adoption," *arXiv preprint 2308.12258,* 2023.

[9] N. Humble, "Risk management strategy for GenAI in computing education: how to handle the strengths, weaknesses, opportunities, and threats?," *International Journal of Educational Technology in Higher Education,* vol. 21, 61, 2024.

[10] C. Sengul, R. Neykova, and G. Destefanis, "Software engineering education in the era of conversational AI: current trends and future directions," *Frontiers in Artificial Intelligence,* vol. 7:1436350, 2024.

[11] V. D. Kirova, C. S. Ku, J. R. Laracy, and T. J. Marlowe, "Software engineering education must adapt and evolve for an LLM environment," in *Proceedings of the ACM Technical Symposium on Computer Science Education V. 1*, 2024, pp. 666–672.

[12] H. Keuning *et al.*, "Students' Perceptions and Use of GenAI Tools for Programming Across Different Computing Courses," in *Proceedings of the Koli-Calling International Conference on Computing Education Research*, 2024, pp. 1–12.

[13] Y. Jia, C. Li, and A. Chan, "Human-Al Collaboration in the STEM Classroom: A Systematic Literature Review of GenAI as a Complement in Higher Education," in *International Conference on Technology in Education*, 2025, pp. 389–401.

[14] L. Stienstra, A. Mohamed, and M. Mohamed, "Exploring GenAI as a Tutoring Tool: A Case Study in First-Year Computer Programming," in *Proceedings of the 30th ACM Conference on Innovation and Technology in Computer Science Education V. 1*, 2025, pp. 452–457.

[15] M. Nyaaba *et al.*, "Generative AI as a Learning Buddy and Teaching Assistant: Pre-service Teachers' Uses and Attitudes," *arXiv preprint arXiv:2407.11983,* 2024.

[16] M. Leon, "Leveraging Generative AI for On-Demand Tutoring as a New Paradigm in Education," *International Journal on Cybernetics & Informatics,* vol. 13, no. 5, pp. 17–29, 2024.

[17] H. Bai, W. C. Lui, and P. V. Khiatani, "Promoting student engagement with GPTutor: An intelligent tutoring system powered by generative AI," *International Journal of Educational Technology in Higher Education,* vol. 22, no. 1, p. 77, 2025.

[18] M. Perkins, L. Furze, J. Roe, and J. MacVaugh, "The Artificial Intelligence Assessment Scale (AIAS): A framework for ethical integration of generative AI in educational assessment," *Journal of University Teaching and Learning Practice,* vol. 21, no. 6, pp. 49–66, 2024.

[19] UK's Joint Information Systems Committee (JISC), "AI maturity toolkit for tertiary education," in *www.jisc.ac.uk/ai-maturity-toolkit-for-tertiary-education*, Last accessed: Jan. 2026.

[20] M. Waqas, A. Hania, and X. U. Chunyan, "Understanding AIgiarism in higher education: the lens of general AI attitudes and moral disengagement," *Studies in Higher Education,* pp. 1–17, 2025.

[21] C. K. Y. Chan, "Students' perceptions of 'AI-giarism': investigating changes in understandings of academic misconduct," *Education and Information Technologies,* vol. 30, no. 6, pp. 8087–8108, 2025.

[22] A. Bandura and R. H. Walters, *Social learning theory*. Prentice hall Englewood Cliffs, NJ, 1977.

[23] F. Ouyang and P. Jiao, "Artificial intelligence in education: The three paradigms," *Computers and Education: Artificial Intelligence*, vol. 2, p. 100020, 2021.

[24] D. Gibson, V. Kovanovic, D. Ifenthaler, S. Dexter, and S. Feng, "Learning theories for artificial intelligence promoting learning processes," *British Journal of Educational Technology,* vol. 54, no. 5, pp. 1125–1146, 2023.

[25] S. Rashid, "Habit Predicting Higher Education EFL Students' Intention and Use of AI: A Nexus of UTAUT-2 Model and Metacognition Theory," *Education Sciences,* vol. 15, no. 6, p. 756, 2025.

[26] V. Garousi, Z. Jafarov, A. Movsumova, A. Namazov, and H. Mirzayev, "Encouraging Responsible GenAI Use in Software Engineering Education: A Design-Oriented Model," *In Press, Journal of Systems and Software,* 2026.

[27] V. Garousi. "Repository of Software Testing Laboratory Courseware." https://bit.ly/garousi-openware-software-testing-labs (accessed June 2025.

[28] V. Garousi, "An Open Modern Software Testing Laboratory Courseware: An Experience Report " in *Proceedings of the IEEE Conference on Software Engineering Education and Training*, 2010, pp. 177–184.

[29] J. R. Young, D. Bevan, and M. Sanders, "How Productive Is the Productive Struggle? Lessons Learned from a Scoping Review," *International Journal of Education in Mathematics, Science and Technology,* vol. 12, no. 2, pp. 470–495, 2024.

[30] C. Wohlin, P. Runeson, M. Höst, M. C. Ohlsson, B. Regnell, and A. Wesslén, *Experimentation in Software Engineering: An Introduction*. Kluwer Academic Publishers, 2000.

## APPENDIX A: DESIGN OF THE ANONYMOUS QUESTIONNAIRES

### A.1 – Bloom-Level Learning Support Questionnaire (Students): Addressed RQ1

**Purpose:**
To explore how students perceived GenAI to support (or not support) their learning at different Bloom cognitive levels.

| Question | Response Type |
|---|---|
| 1. In what ways, if any, did GenAI support your learning at the Remember level (e.g., recalling concepts or terminology)? | Free-text |
| 2. In what ways, if any, did GenAI support your learning at the Understand level (e.g., explaining concepts in your own words)? | Free-text |
| 3. In what ways, if any, did GenAI support your learning at the Apply level (e.g., applying known methods or techniques)? | Free-text |
| 4. In what ways, if any, did GenAI support your learning at the Analyze level (e.g., identifying weaknesses, assumptions, or alternatives)? | Free-text |
| 5. In what ways, if any, did GenAI support your learning at the Evaluate level (e.g., comparing solutions or judging quality)? | Free-text |
| 6. In what ways, if any, did GenAI support your learning at the Create level (e.g., generating ideas based on your own inputs)? | Free-text |
| 7. Please briefly describe how GenAI supported (or did not support) your learning across these cognitive levels. | Free-text |

### A.2 – GenAI Usage Practices Questionnaire (Students): Addressed RQ2

**Purpose:**
To understand how explicit Bloom-level guidance influenced students' GenAI usage practices.

| Question | Response Type |
|---|---|
| 1. Describe when you typically used GenAI during learning activities (e.g., before, during, or after working on a task). | Free-text |
| 2. How did Bloom-level guidance influence *when* you decided to use GenAI? | Free-text |
| 3. How did Bloom-level guidance influence *how* you interacted with GenAI? | Free-text |
| 4. Describe a situation where Bloom-level guidance changed the way you would normally have used GenAI. | Free-text |
| 5. Were there situations where you intentionally avoided using GenAI? If so, why? | Free-text |

### A.3 – Perceived Benefits and Drawbacks Questionnaire (Students): Addressed RQ3 – student perspective

**Purpose:**
To elicit students' perceived benefits and limitations of the Bloom-aligned GenAI approach.

| Question | Response Type |
|---|---|

| | | |
|---|---|---|
| 1. | What aspects of the Bloom-aligned GenAI guidance did you find most beneficial for your learning? | Free-text |
| 2. | What challenges or limitations did you experience when using GenAI under this approach? | Free-text |
| 3. | Did this approach change how you think about using GenAI for learning? Please explain. | Free-text |
| 4. | Would you like this type of GenAI guidance to be used in other courses? Why or why not? | Free-text |

**A.4 – Instructor Reflection Questionnaire (Instructors): Addressed RQ3 – instructor perspective**

**Purpose:**

To capture instructors' reflections on pedagogical value, feasibility, and limitations.

| Question | | Response Type |
|---|---|---|
| 1. | How did explicit Bloom-level GenAI guidance influence students' learning practices and engagement? | Free-text |
| 2. | What pedagogical benefits of this approach did you observe? | Free-text |
| 3. | What challenges or drawbacks did you encounter when implementing this approach? | Free-text |
| 4. | How did this approach affect your own teaching practices or assessment design? | Free-text |
| 5. | Would you adopt or adapt this approach in future courses? Please explain. | Free-text |

**A.5 – Student Background and GenAI Familiarity Questionnaire**

**Purpose:**

To contextualize student responses by capturing prior GenAI experience.

| Question | | Response Type |
|---|---|---|
| 1. | Before this course, how had you used GenAI tools for learning, if at all? | Free-text |
| 2. | How confident did you feel using GenAI tools at the start of the course? Please explain. | Free-text |
| 3. | Had you previously received any guidance on responsible or structured GenAI use? If so, describe it briefly. | Free-text |